\documentclass{aa}  

\usepackage{graphicx}
\usepackage{txfonts}
\usepackage{lscape}
\usepackage{longtable}

\newcommand{\feii}{[\ion{Fe}{ii}]}
\newcommand{\sii}{[\ion{S}{ii}]}

\newcommand{\oii}{[\ion{O}{ii}]}
\newcommand{\oi}{[\ion{O}{i}]}

\newcommand{\nii}{[\ion{N}{ii}]}

\newcommand{\osei}{[\ion{O}{i}]6300\AA}

\newcommand{\kms}{km\,s$^{-1}$}
\newcommand{\um}{$\mu$m}

\newcommand{\lsun}{L$_{\odot}$}
\newcommand{\msun}{M$_{\odot}$}
\newcommand{\msunyr}{M$_{\odot}$\,yr$^{-1}$}
\newcommand{\macc}{$\dot{M}_{acc}$}
\newcommand{\mloss}{$\dot{M}_{jet}$}

\newcommand{\lacc}{$L_{\mathrm{acc}}$}

\newcommand{\lstar}{$L_{\mathrm{*}}$}
\newcommand{\mstar}{$M_{\mathrm{*}}$}

\newcommand{\cmt}{cm$^{-3}$}

\newcommand{\llvc}{$L_{[\ion{O}{i}]\rm{LVC}}$}
\newcommand{\lhvc}{$L_{[\ion{O}{i}]\rm{HVC}}$}
\newcommand{\loi}{$L_{[\ion{O}{i}]}$}
\newcommand{\sig}{$\sigma$}

\begin{document}

   \title{Connection between jets, winds and accretion in T Tauri stars}

   \subtitle{The X-shooter view\thanks{Based on Observations collected with X-shooter at the Very Large Telescope on Cerro Paranal (Chile), operated by the European Southern Observatory (ESO). Programme IDs: 084.C-0269, 084.C-1095, 085.C-0238, 085.C-0764, 086.C-0173, 087.C-0244, 089.C-0143, 090.C-0253, 093.C-0506, 094.C-0913, 095.C-0134 and 097.C-0349.}}

   \author{B. Nisini
          \inst{1}
          \and S. Antoniucci
          \inst{1}
          \and J.~M. Alcal\'a
          \inst{2}
          \and T. Giannini
          \inst{1}           
          \and C.~F. Manara
          \inst{3}  
          \and A. Natta
          \inst{4,5} 
          \and D. Fedele
          \inst{4}   
          \and K. Biazzo
          \inst{6}                              
          }

   \institute{INAF - Osservatorio Astronomico di Roma, Via di Frascati 33, 00040 Monte Porzio Catone, Italy\\
              \email{nisini@oa-roma.inaf.it}
         \and
             INAF - Osservatorio Astronomico di Capodimonte, Salita Moiariello 16, 80131, Napoli, Italy
             \and
             Scientific Support Office, Directorate of Science, European Space Research and Technology Centre (ESA/ESTEC), Keplerlaan 1, 2201 AZ Noordwijk, The Netherlands 
             \and
             INAF - Osservatorio Astrofisico of Arcetri, Largo E. Fermi, 5, 50125 Firenze, Italy
             \and DIAS/School of Cosmic Physics, Dublin Institute for Advanced Studies, 31 Fitzwilliams Place, Dublin 2, Ireland
                        \and
             INAF-Osservatorio Astrofisico di Catania, via S. Sofia 78, 95123, Catania, Italy
              }

   \date{}

 
  \abstract{Mass loss from jets and winds is a key ingredient in the evolution
  of accretion discs in young stars. While slow winds have been recently extensively studied
  in T Tauri stars, little investigation has been devoted on the occurrence of high velocity
  jets and on how the two mass-loss phenomena are connected with each other, and with the 
  disc mass accretion rates.
  In this framework, we have analysed the \osei\, line in a sample of 131 young stars with discs in the Lupus, Chamaeleon and $\sigma$ Orionis star forming regions. The stars were observed 
with the X-shooter spectrograph at the Very Large Telescope (VLT) and have mass accretion rates spanning from 10$^{-12}$ to 10$^{-7}$ \msunyr . 
The line profile was deconvolved into a low velocity component (LVC, |V$_r$| $<$ 40\kms) 
and a high velocity component (HVC, |V$_r$| $>$ 40\kms ), originating from slow winds and 
high velocity jets, respectively. The LVC is by far the most frequent component, with a detection rate
of 77\%, while only 30\% of sources have a HVC. 
The fraction of HVC detections slightly increases (i.e. 39\%) in the sub-sample of stronger accretors 
(i.e. with log (\lacc/\lsun)$> -$3). 
The \osei\, luminosity of both the LVC and HVC, when detected, correlates with 
stellar and accretion parameters of the central sources (i.e. \lstar , \mstar , \lacc , \macc),
with similar slopes for the two components. The line luminosity correlates better (i.e. has a lower dispersion) 
with the accretion luminosity than with the stellar luminosity or stellar mass.
We suggest that accretion is the main drivers for the line excitation
and that MHD disc-winds are at the origin of both components. In the sub-sample of Lupus sources observed with ALMA a relationship is found 
between the HVC peak velocity and the outer disc inclination angle,
as expected if the HVC traces jets ejected perpendicularly to the disc plane.
Mass ejection rates (\mloss) measured from the detected HVC \osei\, line luminosity span from $\sim$ 10$^{-13}$
to $\sim$ 10$^{-7}$ \msunyr. 
The corresponding \mloss/\macc\, ratio ranges from $\sim$0.01 to $\sim$0.5, with an average
value of 0.07. However, considering the upper limits on the HVC, we infer 
a \mloss/\macc\, ratio $<$ 0.03 in more than 40\% of sources. We argue that
most of these sources might lack the physical conditions needed
for an efficient magneto-centrifugal acceleration in the star-disc interaction region.
Systematic observations of populations of younger stars, that is, class 0/I, 
are needed to explore how the frequency and role of jets evolve during the pre-main sequence 
phase. This will be possible in the near future thanks to space facilities such as the James Webb space telescope (JWST). 
   }


   \maketitle
%

\section{Introduction}

The evolution and dispersal of circumstellar discs is an important step in the star and 
planet formation process. disc dissipation in the first stages of pre-main sequence (PMS) evolution
occurs due to the simultaneous effects 
of mass accretion onto the star and the ejection of matter from the star-disc system.
Mass loss, in particular, follows different routes, that involve stellar winds, slow disc-winds,
both photo-evaporated and magnetically driven,
and high velocity and collimated jets.  The latter are the most spectacular 
manifestation of mass loss phenomena, as they often extend over large spatial scales 
interacting with the ambient medium and originating the well known chains of 
shock spots known as Herbig-Haro objects (e.g. Bally 2016 and references therein).
It is widely accepted that proto-stellar jets originate from the surface of accretion
discs via a magneto-centrifugal launching mechanism. 
They are observed over the entire PMS life of a solar type star, 
from the embedded class 0 phase to the class II phase (Frank et al. 2014). 

It is indeed a common paradigm that jets are a ubiquitous phenomenon during the 
active phase of accretion, as they are able to efficiently extract angular momentum 
accumulated in the disc, allowing matter in the disc to accrete onto the star. 
Millimetre observations of class 0 sources confirm that 
bipolar outflows and jets are ubiquitous at this evolutionary stage and 
represent a fundamental output of the star formation process 
(e.g. Mottram et al. 2017, Nisini et al. 2015).
The role of jets in more evolved sources is less clear, although powerful optical
jets are commonly observed in very active Classical T Tauri (CTT) stars. Due to the low extinction,
they have been studied in details employing diagnostics based on line emission in a wide
spectral range, from X-ray to IR wavelength (see e.g. Frank et al. 2014 and references therein).
In few cases there is evidence that
jets can efficiently extract the angular momentum, confirming their importance
in mediating the accretion of matter from the disc onto the star (e.g. Bacciotti et al. 2002, Coffey et al. 2007).
However, the frequency of the jet phenomenon among the general population of class II stars
is still far from being settled. 
Searches for shocks and
outflows through deep optical and IR imaging have been performed in several nearby star-forming regions,
such as Lupus and Chamaeleon, and they have revealed the
presence of only few tens of distinct flows (e.g. Bally et al. 2006, Wang \& Henning 2009) 
against populations of more than 200 young stellar objects (Luhman 2008, Evans 2009).
This number represents however a lower limit on the jet occurrence in T Tauri
stars, as for most sources the presence of a small-scale micro-jet could escape
detection in large surveys with poor spatial resolution.
The question on how common jets are in active accreting stars has a major relevance in the context
of disc evolution, as jets can significantly modify 
the structure of the disc region involved in the launching with respect to a standard model that assumes only 
turbulence as the main agent for angular momentum dissipation (e.g. Combet \& Ferreira 2008).
In addition, jets can shield stellar radiation from reaching the disc surface,
again altering the properties of the upper disc layers. 

A direct way to infer the presence of high velocity gas in poorly embedded young stars
 is through observations of optical atomic forbidden lines, which are excited in the dense and warm
shocked gas along the jet stream. Several surveys of forbidden 
lines in CTT stars have been so far conducted, mostly directed towards the bright
sources of the Taurus-Auriga star forming region 
(e.g. Hamann 1994, Hartigan et al. 1995, henceforth H95, Hirth et al. 1997, Antoniucci et al. 2011, 
Rigliaco et al. 2013,
 Simon et al. 2016), and recently on a sample of sources in the Lupus and $\sigma$ Ori regions
(Natta et al. 2014, henceforth N14).
Such surveys have in particular shown that the \osei\, transition is the brightest 
among the forbidden lines due to the combination of oxygen large abundance and favourable excitation 
conditions in the low ionization environment of the outflow. 

The \osei\, line commonly exhibits two distinct components, as originally recognized
in the pioneering work of Edwards et al. (1987): a low 
velocity component (LVC), peaking at velocities close to systemic one or only slightly
blue-shifted, and a high velocity component (HVC) observed at velocities up to $\sim$ $\pm$ 200\kms . 
The origin of the LVC is still unclear, but it is quite well established that
it comes from a distinct and denser region with respect to the HVC (H95, N14).
The fact that the LVC is often blueshifted by up to 30\kms indicates that it is 
associated with an atomic slow disc-wind, although a contribution from 
bound gas in Keplerian rotation in the inner disc has been suggested by 
recent observations at higher resolution (Rigliaco et al. 2013,
Simon et al. 2016). 
The two main mechanisms invoked to explain the formation of slow disc-winds are 
photoevaporation of the gas in the disc by means of 
EUV/X-ray stellar and/or accretion photons (the so-called photoevaporative winds, Ercolano \& Owe 2010, 2016; Gorti et al. 2009) or winds driven by the magnetic field threading the disc (the 
magheto-hydro-dynamic, MHD, disc-winds, i.e. Casse \& Ferreira 2000, Bai et al. 2016). 
As photoevaporation has been recognized to be an important mechanism for
disc dispersal in late stages of pre-main sequence evolution (e.g. Alexander et al. 2014), 
great efforts have been given to model line excitation and profiles of the LVC in this framework (Ercolano \& Owen 2010, 2016). 

The HVC on the other hand can be directly connected with the extended collimated jets, as shown 
by long-slit spectroscopy performed along the flow axis in known jet driving sources.  
H95 was the first to use the HVC \osei\, luminosity in CTTs of the Taurus-Auriga clouds
 to measure the jet mass flux rate (\mloss)
and estimate its ratio with the mass accretion rate (\mloss/\macc\, ratio) in a sample of 22 sources. This is a fundamental
parameter that characterizes the jet efficiency in regulating the 
mass transfer from the disc to the star. 
Subsequent studies have revised the \macc\, determinations on some of the H95 sources
 adopting different methods (e.g. Gullbring et al. 1998, Muzerolle et al. 1998), suggesting that the 
 \mloss/\macc\, ratio in the sample of Taurus sources spans 
 between $\sim$ 0.05 and 1. Determinations of this ratio as derived
from detailed studies of individual jets agree with these values 
in a wide range of masses, and for young and embedded sources 
(Cabrit 2007, Antoniucci et al. 2008, Antoniucci et al. 2016).
All these estimates suffer, however, from the 
non-simultaneous determination of the accretion and ejection parameters, 
or from different adopted methods and assumptions.
Consequently, it is not clear whether the large scatter in the derived ratio is due to
the non-homogeneity of the approach or to a real effect.
An adequate answer to this problem can be given only 
by applying the same methodology 
to a statistically significant sample of sources where accretion and ejection tracers are
simultaneously observed. 

In this paper, we present an analysis conducted on a sample of 131 class II sources 
(mostly CTT stars) in the Lupus, Chamaeleon and $\sigma$ Orionis star-forming regions,
whose UV and optical spectra have been acquired with the X-shooter instrument
  (Vernet et al. 2011) at the Very Large Telescope (VLT). 
  These observations are part of a project aimed at 
  characterizing the population of young stellar objects with discs in nearby star-forming regions
  (Alcal\'a et al. 2011).
As part of this project, the stellar and accretion parameters of the observed stars 
have been consistently and uniformly derived (Rigliaco et al. 2012, Alcal\'a et al. 2014, 2017, Manara et al. 2016a, 2017, Frasca et al. 2017). The same sample of Lupus observed with X-shooter has been 
also observed with ALMA (Ansdell et al. 2016) and the inferred disc masses were found to be correlated with 
the mass accretion rates (Manara et al. 2016b). For a sub-sample of these sources, the 
HI line decrements have been analysed and the general properties of the hydrogen emitting
gas derived (Antoniucci et al. 2017).

A study of the forbidden line emission in a sub-sample of 44 sources in Lupus and $\sigma$ Orionis
is presented in N14. These authors concentrated their analysis on the properties of the LVC
and derived correlations between the LVC luminosity and accretion and stellar
luminosities. In addition, ratios of different lines showed that
the LVC originates from a slow dense wind ($n_e > 10^8$\cmt) which is mostly neutral. 
Here we extend the work of N14 to a much larger sample of objects covering
a wider range of stellar and accretion properties.
Our work is, however mainly focussed on the HVC and aims
at investigating the occurrence of jets during the class II phase,
measuring the jet efficiency and studying whether it varies with the source properties. 
In addition, relationships between the LVC and HVC will be also discussed, 
with the goal of defining any connection between the phenomena giving rise to the two emissions. 

The paper is organized as follows: Section 2 presents the data sample adopted for our analysis.
In Section 3 we describe the procedure adopted to retrieve the parameters of
the LVC and HVC from the \osei\, line profiles, while Section 4 summarizes the statistics
on the detection of the two components. In Sections 5 and 6 correlations between the \osei\,
luminosity and stellar and accretion parameters are derived and the \mloss/\macc\, ratio 
discussed. Section 7 describes kinematical information on the HVC and LVC derived from 
line profiles. Discussion and conclusions are finally presented in Sections 8 and 9, respectively.

\section{The data sample }

The adopted sample consists of 131 Young Stellar Objects (YSOs) observed with the X-shooter instrument. 
In particular, the sample includes sources 
in the Lupus, Chamaeleon and \sig\, Orionis star forming regions, whose X-shooter spectra have been already analysed in
previous papers (Alcal\'a et al. 2014 and 2017 for Lup; Rigliaco et al. 2012 for $\sigma$ Ori; 
Manara et al. 2016a and Manara et al. 2017 for Cha). 
Details on observations and data reduction are given in those papers
 while Table 1 summarizes the instrument setting for each sub-sample of sources.
 
The Lupus sources comprise 82 confirmed YSOs of the Lupus I, II, III, IV clouds, and constitute the most complete and homogeneous sub-sample analysed in this work. They, in fact, represent about 90\% of all the candidate class II in these clouds. Thirty-six of these sources have been observed as
part of the INAF/GTO X-shooter programme while the remaining 40 are part of two individual 
follow-up programmes (095.C-0134 and 097.C-0349, P.I. J. Alcal\'a). The two samples were observed with the same 
X-shooter setting, but the spectral resolution of the more recently acquired sample is slightly lower 
($\sim$7500 vs. 8800 in the VIS arm) due to a change of the instrument slit wheel between 
the two sets of observations. We have also included ESO archival data for six additional sources (see Alcal\'a et al. 2017), 
that have been observed with a narrower slit at a higher resolution ($\sim$ 18\,000).

The Chamaeleon sample comprises 25 of the accreting sources analysed in Manara et al. (2016a)
 plus 16 sources observed in the programme 090.C-0253
(P.I. Antoniucci) for a total of 41 objects (31 in Cha I and 10 in Cha II). These observations
were all performed at the highest X-shooter resolution (see Table 1). 
All the considered Cha I sources are part of the larger sample presented in Manara et al. (2017)
while the Cha II objects are presented here for the first time. The derived stellar and
accretion parameters for these additional sources are discussed in Appendix A and summarized in Table A.1. Finally, we have also included in the sample the eight sources in the \sig\, Ori cloud studied
 in Rigliaco et al. (2012) and Natta et al. (2014).

The stellar and accretion parameters (namely \lstar , extinction, spectral type, 
\lacc) of all stars have been self consistently
derived by fitting the X-shooter spectrum with the sum of a stellar photosphere and an excess
continuum due to accretion (see Alcal\'a et al. 2017 and Manara et al. 2016a for details). Stellar mass depends on the adopted evolutionary tracks, and we adopt here masses derived from the Baraffe et al. (2015) tracks for all samples,
with the exception of the \sig\, Ori objects whose masses have been derived by Rigliaco et al. (2012)
from the Baraffe et al. (1998) tracks.
Accretion luminosities were eventually converted into mass accretion rates (\macc)
using the relation \macc $\approx 1.25 ~ \frac{L_{acc} R_{\star}}{G M_{\star}}$,
where $R_{\star}$, the stellar radius, was calculated from the effective temperature and 
stellar luminosity (for more details see, e.g. Alcal\,a' et al. 2014 ).

The list of considered sources is given in Table B.1. 
Given the slightly different setting for the Lupus observations, we maintained the division
in three different sub-samples (GTO, New sample, Archive) as done in Alcal\'a et al. (2017).
For convenience, the table also reports all the relevant parameters collected from the individual papers.
Considering the global sample, masses range between 0.02 and 2.13 \msun, log (\lacc/\lsun) between $-$5.2 and $+$0.85,
and the corresponding mass 
accretion rates are between $\sim$ 10$^{-12}$ and 10$^{-7}$ \msunyr , meaning that they span approximately
five orders of magnitude. 

The majority of the sources of our global sample are class II CTTs. The sample also includes a small group (12) of transitional discs (TD) in Lupus,  that is, T Tauri sources having discs with large inner holes depleted of dust. 
In addition, the sample includes 14 sources
with mass $\la$ 0.1 \msun , thus comprising possible brown dwarf (BD) candidates. We will
discuss the properties of these stars separately in the following analysis as their accretion
and ejection properties might be different from those of CTT stars.

Finally, on the basis of the derived stellar parameters, seven sources in Lupus have
been classified as 'sub-luminous', that is they have a stellar luminosity much lower than 
that of the other sources of the cloud with the same spectral type. They are likely to be sources with
the disc seen edge-on, hence partially scattering the stellar continuum. Two of these sources
(namely Sz133 and Sz102 in Lupus) fall below the zero-age main sequence and thus an estimate
of the mass (and mass accretion rate) was not obtained (Alcal\'a et al. 2017).
Finally, six sources in Lupus have an accretion rate compatible with chromospheric emission and thus
their $L_{acc}$ value has been considered here as an upper limit. 
All the above cases are indicated in the notes of Table B.1.

\begin{footnotesize}

\begin{table*}
\caption[]{Details on the used X-shooter samples}
\vspace{0.5cm}
\begin{tabular}{lccccc}
\hline
Sample & N. & X-shooter slit &  Resolution & Ref. parameters & Ref. \oi\, fluxes \\
\hline
Lupus GTO& 36 &1\farcs 0/0\farcs 9/0\farcs 9 & 5100/8800/5600 & Alcal\'a et al. (2014,2017) & Natta et al. 2014 \\
                 &  &                             & 59/34/54 \kms & \\
Lupus New Sample& 40 &1\farcs 0/0\farcs 9/0\farcs 9 & 4350/7450/5300 & Alcal\'a et al. (2017) & this work \\
                 &  &                             & 69/40/57 \kms   &    \\
Lupus archive & 6   &0\farcs 5/0\farcs 4/0\farcs 4 & 9900/18200/10500 & Alcal\'a et al. (2017) & this work \\
                 &  &                             & 69/16.5/29 \kms  &  &   \\
Cha I & 31 &0\farcs 5/0\farcs 4/0\farcs 4 & 4350/18200/10500 & Manara et al. (2016a,2017) & this work  \\
                 &  &                             & 69/16.5/29   &  &   \\
Cha II& 10 &0\farcs 5/0\farcs 4/0\farcs 4 & 9900/18200/10500 & this work, Manara et al. (2017) & this work \\
                 &  &                             & 69/16.5/29   &  &   \\          
$\sigma$ Ori & 8 &1\farcs 0/0\farcs 9/0\farcs 9 & 5100/8800/5600 & Rigliaco et al. 2012 & Natta et al. 2014 \\  
                 &  &                             & 59/34/54 \kms &  \\

\hline\\[-5pt]
\end{tabular}
\end{table*} 
\end{footnotesize}

\section{ \oi\, profile decomposition}

\begin{figure*}
\includegraphics[angle=0,width=15cm]{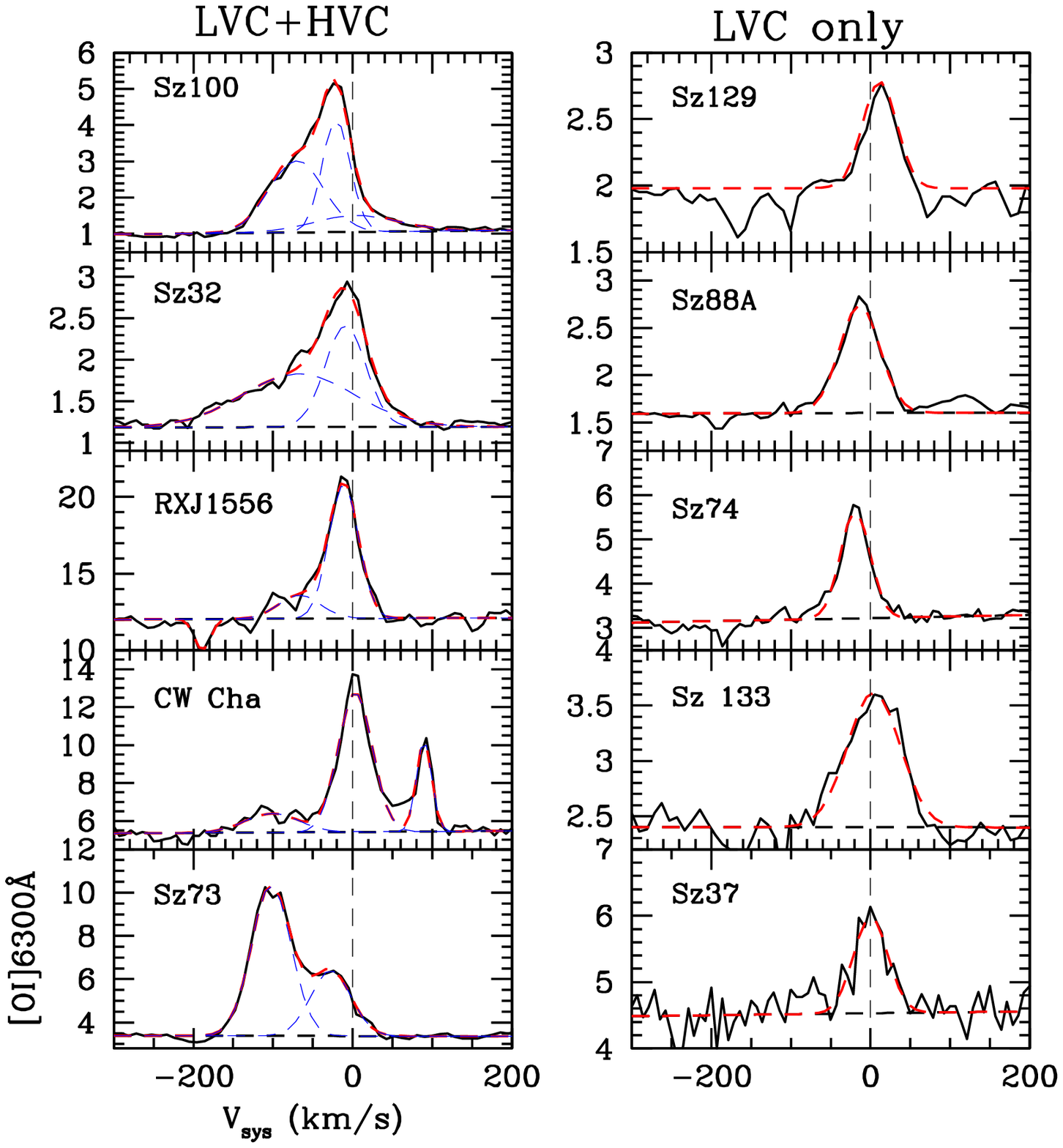}
     \caption{Examples of \osei\, spectra of sources with and without a HVC. Blue-dashed lines 
     indicate the Gaussian profiles used to fit the different components, and the red-dashed line
     is the final best fit.}
         \label{profile}
\end{figure*}   

We focussed our analysis on the \osei\, line which is the brightest
among the class II forbidden lines in the spectral range of our data (e.g. H95, N14). 
The \osei\, emission of the Lupus stars observed in the GTO and \sig\ Ori sample has been analysed
and discussed in N14 together with the emission from other forbidden lines.
We have applied the same methodology as in N14 to the rest of the sample in order
to have an homogeneous database of \oi\, luminosities in the different kinematical
components.

As discussed in the introduction, the \osei\, profile can be decomposed in
a LVC, close to systemic velocity, and a blue- or red-shifted HVC.  
To separately derive line center and width of the different components 
we fitted the \oi\, line profile with one or more Gaussians.
We have used the photospheric rest velocity as measured
from the $\ion{Li}{i}$ line at 6707.876\AA\, to set the zero velocity scale of all spectra.
For the few cases in which the $\ion{Li}{i}$ line was not detected, we used the $\ion{K}{i}$
photospheric line at 7698.974\AA .

Given the limited spectral resolution, we did not attempt fits with more
than three Gaussians (one for the LVC and two for the blue- and red-shifted components of the 
HVC). To be consistent with N14, we defined the velocity that separates the LVC from the HVC as 40 \kms, 
which roughly corresponds to the resolution attained for most of the Lupus objects.
However, the resolution of the observations in Chamaeleon is higher by approximately a factor of two, 
so we are potentially able to identify HVC even if at lower radial velocities. For this sample
we consistently defined any emission with peak radial velocity  |V$_r$| $<$ 40 \kms\, 
 as an LVC if a single component is detected. However, if we detected two components, the one at the 
higher velocity was defined as HVC even if it has a |V$_r$| $<$ 40 \kms . 

We remark that this second component could not be associated to a jet, but rather
to the broad, slightly blue-shifted LVC component with |V$_r$| typically of $\la$~30\kms 
sometimes detected in high resolution observations (e.g. Rigliaco et al. 2013, Simon et al. 2016). Since there are only four of such cases in our sample,
this distinction is not going to affect our statistics.

To perform the deconvolution, we have used the {\sc onedspec} task within the \textit{IRAF}\footnote{IRAF is distributed by the National Optical Astronomy Observatory, which is operated by the Association of Universities for Research in Astronomy (AURA) under a cooperative agreement with the National Science Foundation.} package. The HVC component can be identified as an individual line (often partially blended with the LVC) 
 or as an extended (mostly blue-shifted) wing of the LVC (see Fig. \ref{profile} for some examples). 
 In this latter case, as the result of the deconvolution might be more dependent
 on the initial guess for the Gaussian parameters, we have checked the IRAF  
  deconvolution also using the curve-fitting programme PAN (Peak-analysis) within DAVE (Data Analysis and Visualization Environment),
  running on IDL (Azuah et al. 2009).
 This programme allows one to define the initial guess not only on the velocity peak but also
 on the width of the Gaussians, as well as to have better visualization of the individual
 fitted curves. We find in all cases that a different choice of the initial parameters 
 influences the final deconvolved line flux by no more than 30\% . The largest uncertainty on the peak
 velocities is estimated to be about 5 and 10\kms\, for the LVC and the HVC, respectively. 
When the HVC is not detected, a 3$\sigma$ upper limit is derived from the \textit{rms} of the continuum
computed in regions adjacent to the LVC, multiplied for the instrument resolution element. 

Line fluxes have been corrected for extinction and then converted into intrinsic luminosities 
adopting the A$_V$ and distances values given in Table B.1. All the derived line parameters are listed in Table B.2, where we report, in addition to the new determinations, also the \osei\, line parameters relative 
to the Lupus and \sig\, Ori GTO sample published in N14.

The detection frequency of other forbidden lines tracing the HVC (e.g.\sii\ 6716.4, 6730.8\AA\,
and \nii\ 6548.0, 6583.4\AA ) is always lower than that of the \osei\, line, and their signal-to-noise ratio (S/N) usually worse, therefore we do not discuss them in this paper. However,
for sources where the \sii\ and \nii\ lines are detected with a good enough S/N  
we have performed the Gaussian decomposition into a LVC and HVC and checked that the
resulting kinematical parameters are consistent with those of the \osei\ lines. 
Appendix C summarizes the peak velocities and FWHM derived in these cases, which 
generally agree with those derived for the \osei\, line within $\sim \pm$20 \kms\..

 We have also examined the X-shooter 2D spectral images to see if spatially extended \osei\ emission can be resolved in sources where the HVC component is detected, due to a fortuitous alignment of the instrumental slit with the jet axis. We found that in five of the brightest \osei\ sources the HVC emission peak is shifted with respect to the source continuum by 
$\sim$ 0\farcs 2-0\farcs 4. The position velocity (PV) diagrams relative to these 
sources are presented in Appendix D. In one of the sources, Sz83, the HVC appears
composed by at least four different emission peaks at different velocities, showing
an increase in the spatial displacement with increasing speed. This jet  
acceleration pattern was already found and studied with spectro-astrometric techniques
by Takami et al. (2001). Finally, one of the sources with strong HVC emission, Sz102, 
is a well known driver of an optical jet (Krautter 1986). We do not detect 
any appreciable shift in its emission as the used X-shooter slit was aligned perpendicular
to the jet axis.

\section{Statistics}

The \osei\, line is detected in 101 out of the 131 sources of our sample (i.e. 77\% rate of detection). 
In all detections the LVC is identified, while the HVC is observed in 39 of the 131 objects (i.e. 30\% rate of detection).
These detection rates are similar to what is found in N14 for the authors' limited sample of 44 sources
(84\% and 27\% detections for the \osei\, LVC and HVC, respectively). 
The \lhvc\, spans a wide range of values - between 0.01 and 3 \lsun\ - for the detections.
In the majority of the sources the \lhvc\, is typically fainter than the \llvc\, 
by a factor of two or three, which partially
explains the higher number of LVC detections with respect to the HVC. 
However sensitivity alone is not the only reason for a difference in
detection rate between the LVC and HVC, as the S/N in the LVC line is often larger 
than 10. 
In some of the sources orientation effects could misclassify emission at
high velocity with a LVC. Assuming a total jet maximum velocity of 200~\kms\, (see Section 7) 
any jet inclined by less than 78 degree with respect to the line of sight would have
a $V_{rad} >$ 40\kms and thus will be detected by us as a HVC. 
The probability that a jet in a randomly oriented 
sample of stars has a line-of-sight inclination angle $< \theta \deg$
is P($< \theta$) = 1-cos($\theta$). For $\theta$= 78$\deg$, the probability is
80\% which means that in 20\% of the sources ($\sim$ 25 sources) a high 
velocity jet could be in principle misinterpreted as a LVC. This number is probably 
lower as observations of Chamaeleon objects were performed at higher resolution, thus
in principle able to better identify HVC emission at lower radial velocity in this sub-sample. 
Even taking this effect into account, we still do not detect the HVC in about 50\% 
of sources potentially with favourable inclination.

Non-detection due to orientation effects might be more important for BD and very low mass objects
as the total jet velocity depends on the stellar mass, hence the jet radial velocity 
in these sources might be comparable to our velocity threshold. Indeed BD jets have V$_r$ typically less than 40\kms\, even for low inclinations with respect to the line of sight (e.g.  Whelan et al. 2014). Among the 14 sources of our sample with \mstar\, less than 0.1 \msun , only two have a detected HVC
(20\% of sources). If we exclude these sources from our statistics, the total detection rate 
increases to 32\%. 

Our detection rates can be compared with those of similar studies performed at higher
spectral resolution, thus potentially less affected by orientation effects.
For example, Simon et al. (2016) observed the \osei\, line in a sample of 33
young stellar objects with a resolution of 6.6 \kms\ detecting HVC emission in 13 sources,
thus with a detection rate of 39\%, similar to our study. 
On the other hand the HVC was detected in 25 out of the 42 sources of H95, 
a detection rate of $\sim$ 60\%. This different detection rate 
could be related, in addition to 
the higher resolution, also to the different properties, mass and mass accretion rates in particular, 
 of the H95 sources in comparison with those of our sample.

\begin{figure}
\includegraphics[angle=0,width=12cm,trim = 1cm 1cm 0 0,clip]{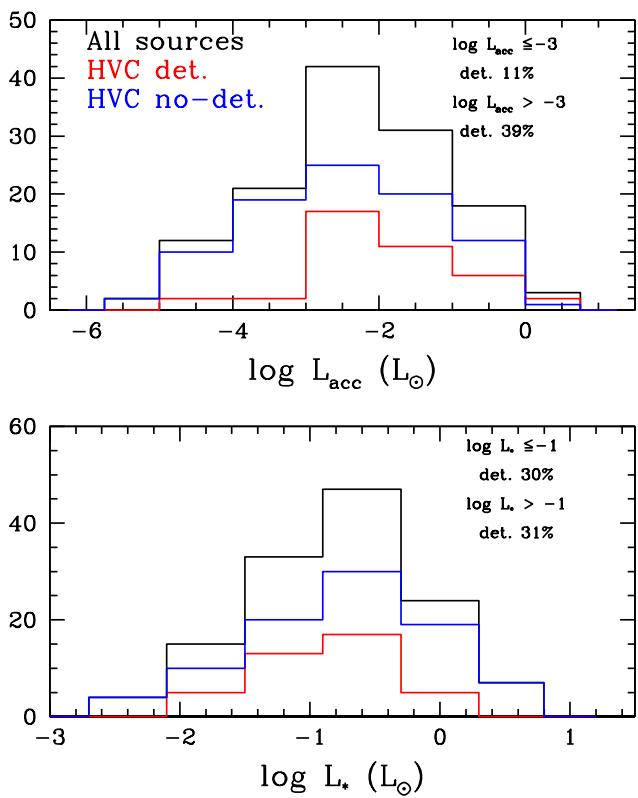}
     \caption{Distribution of accretion (top) and stellar (bottom) luminosity for all
     the sources of our sample is depicted in black.
     Red and blue histograms are relative to sources where the HVC has been detected and not-detected
     respectively. In the upper corner of each plot the percentage of HVC detection 
      in two ranges of luminosities is indicated.}
         \label{histogram}
\end{figure}   

\begin{figure}
\includegraphics[angle=0,width=12cm,trim = 0.8cm 0 0 6cm,clip]{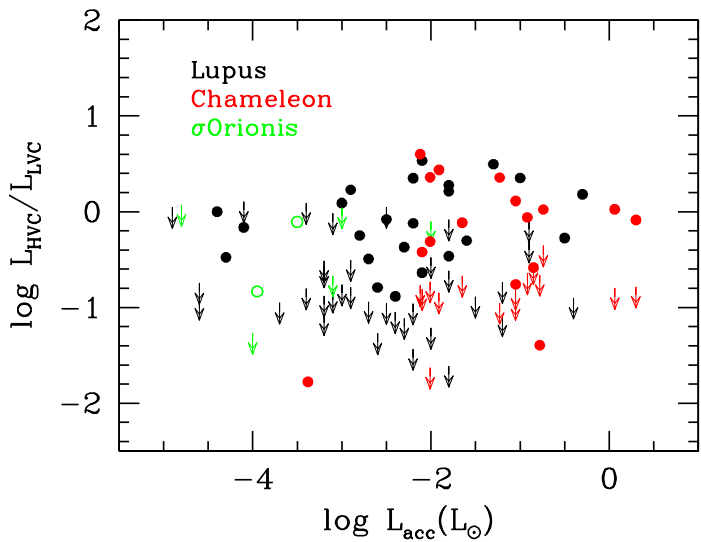}
     \caption{Ratio between the HVC and LVC \osei\, line luminosities as a function of the
     accretion luminosity. Black, red and green symbols indicate sources in Lupus, Chamaeleon 
     and $\sigma$ Orionis clouds, respectively. Arrows refer to 3$\sigma$ upper limits for the HVC luminosity.}
         \label{hvc_lvc}
\end{figure} 

To explore for any dependence of the HVC detection with source parameters, in
Fig. \ref{histogram}
we show the distribution in \lacc\, and \lstar\, separately for sources where the 
HVC has been detected and not-detected. 
It can be seen that
detections are not segregated in specific ranges of parameters.
However, there is a clear trend for higher detection rates in sources with higher \lacc . 
For example, the detection rate decreases from 39\% in sources with 
log (\lacc/\lsun) $> -3$ to 11\% in sources with log (\lacc/\lsun) $\le$ -3.
This trend explains the lower rate of HVC detection in the N14 sample with
respect to our enlarged sample, which contains a larger number of sources with
higher accretion luminosity. No significant variation in the
detection rate is instead observed in sources with different stellar luminosity:
stars with log (\lstar/\lsun) $<$ $-$1 show a similar detection rate as the stars with higher
luminosity.

Figure \ref{hvc_lvc} displays the ratio between the HVC and LVC line luminosity as a function
of the accretion luminosity. When a HVC is detected the luminosity ratio varies between 
$\sim$ 0.1 and 4, with
most of the sources having a \llvc\, luminosity equal or lower than the \lhvc\,. 
No dependence with \lacc\, is found.

Of the 11 TD of our sample all but two have a LVC, while two of them (namely Sz100 and Sz123B)
have also a HVC. These numbers are roughly in line with the overall statistics of our sample (i.e.
80\% LVC, 20\% HVC), and consistent with the fraction of TD with \oi\, LVC in the 
study of Manara et al.(2014). 
This evidence confirms that TD have characteristics similar 
to other CTTs with a full disc, not only for what accretion properties concern (Manara et al. 2014, Espaillat et al. 2014), 
but also regarding the properties of the inner atomic winds. Moreover, the detection of 
the HVC in two of them, in the assumption that this component traces a collimated jet, 
indicates that these sources are undergoing magnetospheric accretion. Finally, of the six sources whose permitted line emission is more
consistent with chromospheric activity rather than with accretion activity, only Lup607 shows \oi\, emission in a LVC. 

\section{Correlation with stellar and accretion parameters}

The \osei\, line luminosity has been correlated with the stellar and accretion parameters
for both the LVC and HVC. Results
are shown in Fig. 4 (\llvc\, vs. $L_{acc}$, $L_{*}$ and $M_{*}$ ), 
Fig. 5 (\lhvc\, vs. $L_{acc}$, $L_{*}$ and $M_{*}$ ), and Fig. 6
(\llvc\, and \lhvc\, vs. \.{M}$_{acc}$). 
The line luminosity correlates well with all the considered parameters in both LVC and HVC
as shown by the Pearson correlation coefficient {\it r} displayed in each plot, which
is always between 0.7 and 0.8, with the exception of the  \lhvc\, vs. $L_{*}$
correlation whose coefficient is 0.63. 
In the \llvc\, vs. $L_{*}$ plot we indicate the locus of the two sub-luminous sources
(namely ParLup 3-4 and Sz102) that significantly deviate from the correlation 
found for the other sources. Given the uncertainty of their stellar luminosity, we
have not included them in deriving the correlation parameters.  
Finally, we have checked that the correlations do not change if we exclude the TD sources from
our sample. 

We performed a best fit linear regression finding the following relationships:
\\
log(\llvc) = (0.59$\pm$0.04)log $L_{acc}$ + (-4.13$\pm$0.11)
\\
log(\llvc) = (1.06$\pm$0.11)log $L_{*}$ + (-4.65$\pm$0.10)
\\
log(\llvc) = (1.87$\pm$0.13)log $M_{*}$ + (-4.52$\pm$0.08)
\\
log(\llvc) = (0.68$\pm$0.05)log $\dot{M}_{acc}$ + (0.69$\pm$0.50)

and
\\
log(\lhvc) = (0.75$\pm$0.08)log $L_{acc}$ + (-4.03$\pm$0.17)
\\
log(\lhvc) = (1.22$\pm$0.21)log $L_{*}$ + (-4.54$\pm$0.18)
\\
log(\lhvc) = (2.13$\pm$0.29)log $M_{*}$ + (-4.48$\pm$0.16)
\\
log(\lhvc) = (0.79$\pm$0.10)log $\dot{M}_{acc}$ + (1.50$\pm$0.86)

We found a slightly flatter slope for the 
\llvc\, vs. $L_{*}$ and \llvc\, vs. $L_{acc}$ correlations with respect to N14 (i.e. 1.37$\pm$0.18 and 0.81$\pm$0.09, respectively).
Our \llvc\, vs. $L_{acc}$ has a slope more similar to that found by Rigliaco et al. (2013)
in a sample of Taurus sources (0.52$\pm$0.07). 
We point out that the present sample 
includes sources with higher \macc\ than the N14 sample alone,
which could be at the origin of the different slope. Correlations with $L_{*}$ present the larger scatter and in general 
we find that both the LVC and HVC show a better correlation with $L_{acc}$ 
than with all the other parameters. 

Mendigut\'ia et al. (2015) investigated the nature of the known correlations between
accretion and emission line luminosities.
 They argue that such correlations cannot be directly attributed 
 to a physical connection between the line formation region and the
 accretion mechanism since they are mainly driven by the fact that both accretion and line luminosity
 correlate with the luminosity of the central source. 
 They suggested to use luminosities normalized by \lstar .
This is what we show in Fig. \ref{corr_lstar}, where the $L_{[\ion{O}{i}]}$/\lstar vs. \lacc/\lstar\, 
correlation is plotted for both the LVC and HVC. We see that \osei\, line to stellar 
luminosity ratio is still correlated with the accretion to stellar luminosity ratio,
supporting the idea that the physical origin of the line is, directly or indirectly,
related to the accretion mechanism. The same conclusion is also suggested by Frasca et al. (2017),
 who found that their derived H$\alpha$ line fluxes 
per unit surface in the sources of the Lupus sample tightly correlate with accretion luminosity,
 excluding any influence from calibration effects.

\begin{figure}[t]
\includegraphics[angle=0,width=15cm,trim = 1cm 0 0 0,clip]{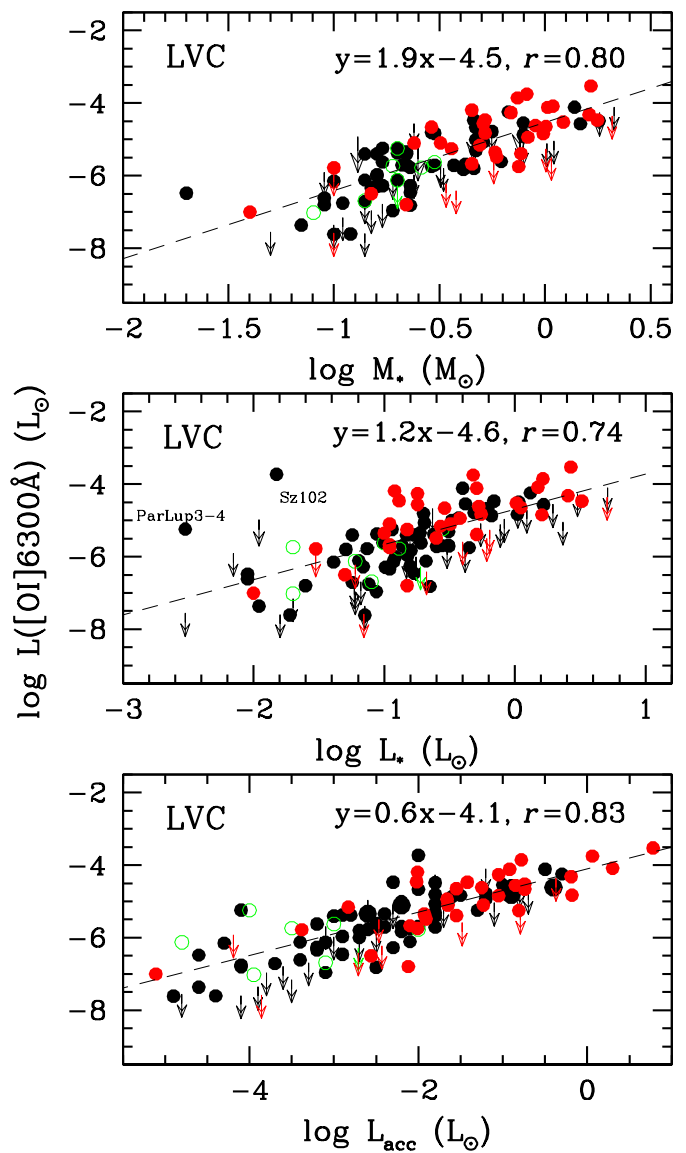}
     \caption{\osei\, line luminosity of the LVC is plotted as a function of, from top to bottom,
     stellar mass, stellar luminosity and accretion luminosity. Colour codes for the 
     symbols are as in Fig. 3. Arrows refer to 3$\sigma$ upper limits.
      The dashed line indicates the linear regression 
     whose parameters are given in the upper right corner of each figure, together
     with the correlation parameter. }
         \label{1}
\end{figure}   

\begin{figure}[t]
\includegraphics[angle=0,width=15cm,trim = 1cm 0 0 0,clip]{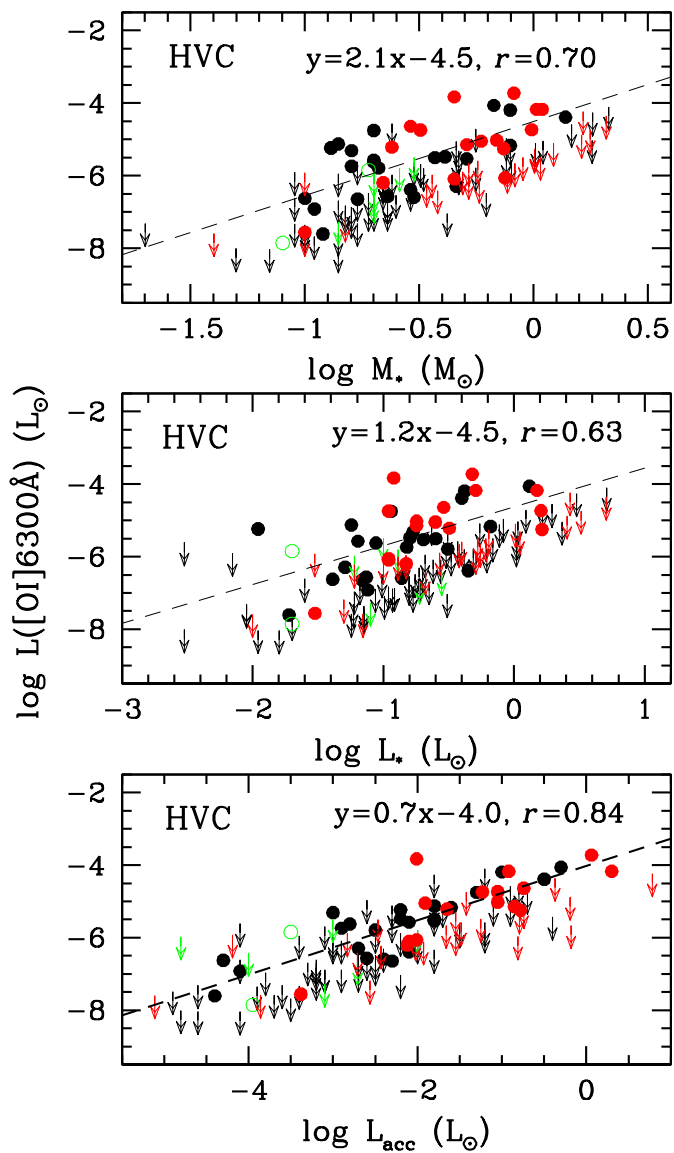}
     \caption{Same as Fig. 4 but for the \osei\, luminosity in the HVC.}
         \label{2}
\end{figure}   

Correlations between \oi\, luminosity and mass accretion rate or accretion luminosity have already been found, and used to derive the mass accretion rate in sources 
where Balmer jump or permitted line luminosities were not detected or uncertain 
(e.g. Herczeg et al. 2008).  This is however the first time that
such correlations are separately found for the LVC and HVC. The log(\lhvc) vs. log \macc\
correlation is particularly useful to indirectly estimate the mass accretion rate in moderately
embedded sources driving jets,  where the permitted lines or the LVC emission region 
might be subject to large extinctions or scattering (e.g. Nisini et al. 2016). 

The derived \llvc\, vs. \lacc\, luminosity relationships can be additionally used to indirectly estimate
the accretion luminosity in the sub-luminous sources. The \lacc\, measured in these
sources from the UV excess could be underestimated due to the fact that part or most
of the UV radiation is scattered away by the edge-on disc. On the contrary, the 
\osei\, emission originates mostly from a more extended region above the disc surface,
thus it is less affected by the disc obscuration (see e.g. Whelan et al. 2014).
Figure \ref{lum_subl} displays the comparison between the two \lacc\, determinations for
the five sub-luminous objects where \osei\, emission has been detected. 
In only two of them, namely Par-Lup 3-4 and Sz102, the \lacc\, derived from \llvc\, 
is significantly larger (up to two order of magnitudes) than the value estimated 
from the UV excess.  We note that extended optical jets have been detected in 
these two sources (Whelan et al. 2014; Krautter 1986), therefore their \osei\, emission 
should be the least contaminated by disc obscuration.

\begin{figure}
\includegraphics[angle=0,width=11cm, trim = 0.7cm 0 0 0,clip]{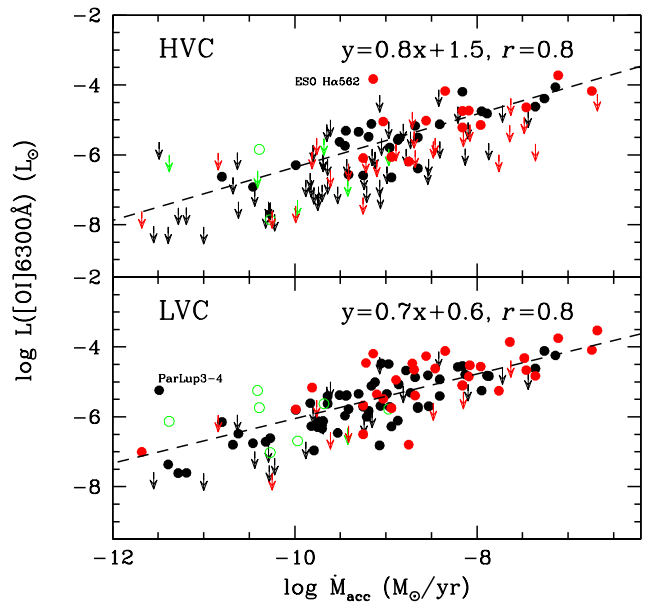}
     \caption{\osei\, line luminosity of the LVC (bottom) and HVC (top) is plotted
     as a function of the mass accretion rate. Symbols and labels are as in Fig. 4.}
         \label{sed}
\end{figure}   

\begin{figure}
\includegraphics[angle=0,width=11cm, trim = 0.7cm 0 0 0,clip]{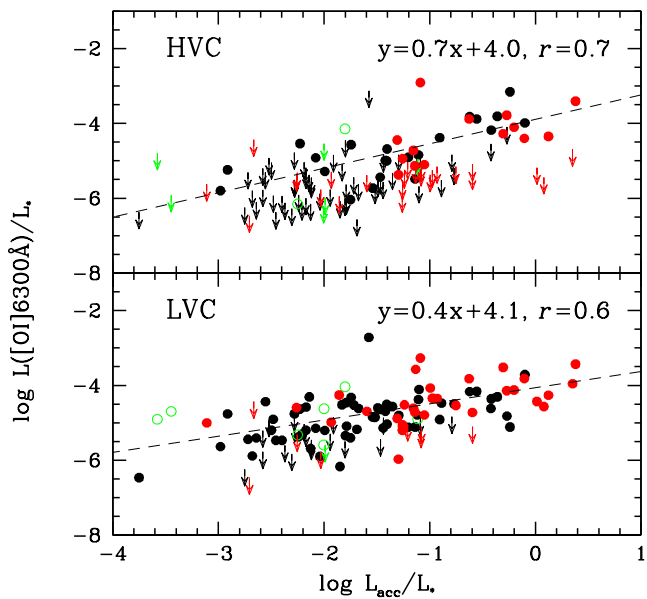}
     \caption{Correlations between the \osei\, line and accretion luminosities, 
     both normalized by the stellar luminosity. Symbols and labels are as in Fig. 4.
     }
         \label{corr_lstar}\
\end{figure}   

\begin{figure}
\includegraphics[angle=0,width=10cm]{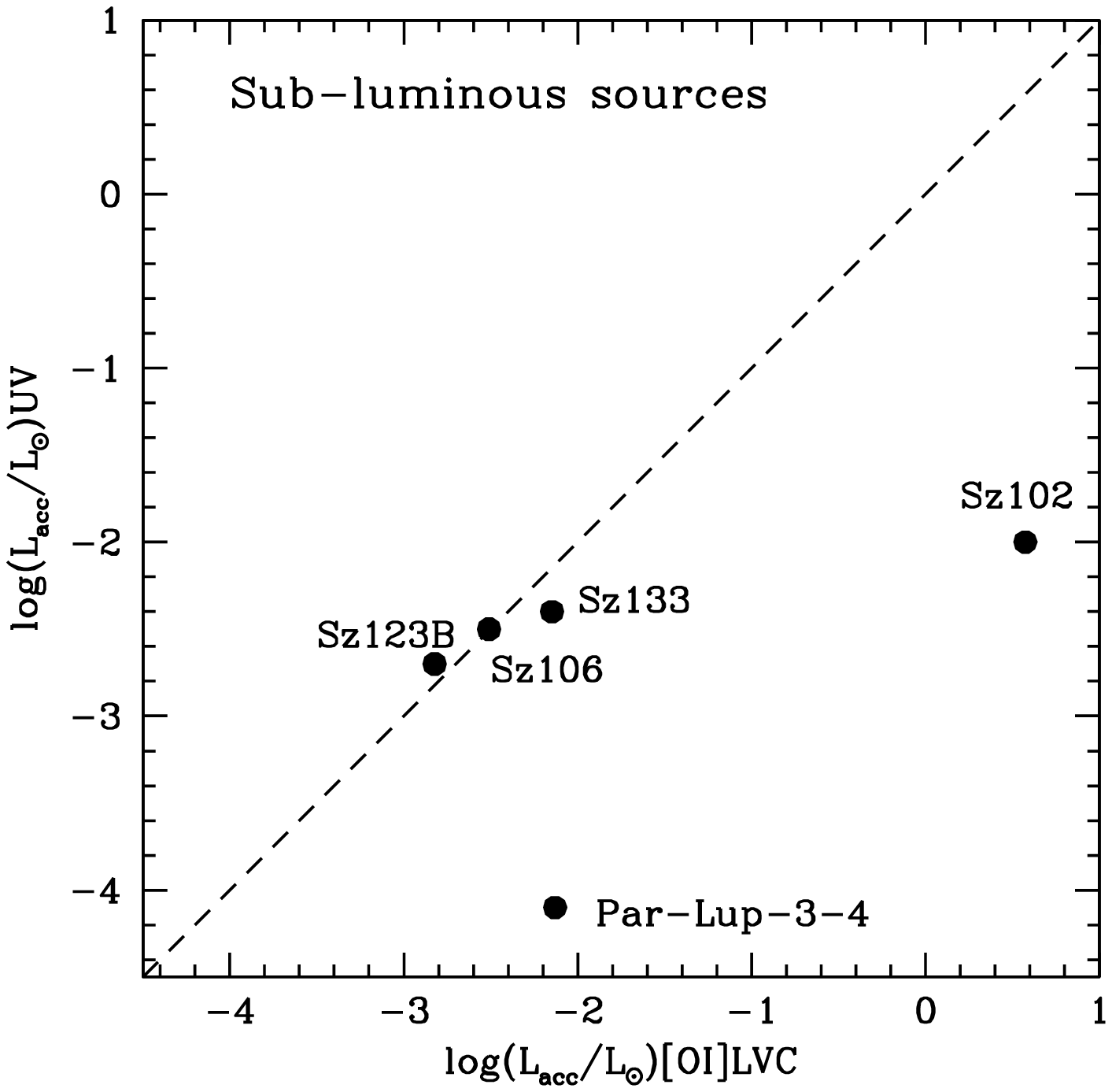}
     \caption{Accretion luminosity derived by modelling the UV excess is compared with that estimated
     through the \llvc\, vs. \lacc\, relationships for the sub-luminous sources.}
         \label{lum_subl}
\end{figure}   

\section{Mass ejection rate and its connection with mass accretion}

In the assumption that the \osei\, HVC traces the high velocity jet, 
its luminosity can be converted into the total mass in the flow (\textit{M})
given that forbidden lines are optically thin. The jet mass flux rate can
be then estimated as $\dot{M}_{jet} = M\times V/l$, where 
$V$ is the flow speed and $l$ is the scale length of the considered
emission. For our calculation we considered $V/l$ = $V_\bot/l_\bot$, where
$V_\bot$ is the component of the velocity projected on the plane of the sky,
and $l_\bot$ is the projected scale length 
of the emission. We took $V_\bot$ to be equal to 100 \kms , assuming a total jet
velocity of 150\kms\, and a median inclination angle of 45$^o$ (see also Section 7),
and  $l_\bot$ equal to the size of the aperture at the given distance (see H95).

To measure the mass of the flow from the line luminosity we have considered the \osei\,
emissivities computed considering a five-level \ion{O}{i} model, and assuming 
$T_e$ = 10\,000 K and $n_e$ = 5$\times$10$^{4}$ cm$^{-3}$. This choice  
is based on the values that typically represent the
region at the base of T Tauri jets, where the \oi\, emission originates.  
In particular, we have considered values estimated from multi-line analysis
that include also the \osei\, line, as in H95, or from 
the analysis of \feii\ lines, that have excitation conditions similar to \oi\, (Giannini
et al. 2013). The adopted values are also similar to those used by H95
to estimate the \mloss\, of a sample of Taurus sources from their HVC \oi\, emission.
Hence, we can make a direct comparison between our derived mass flux rates and those
estimated by H95.
We further assumed that oxygen is neutral with a total abundance of 4.6$\times$10$^{-4}$
(Asplund et al. 2005). This hypothesis is supported by the absence of, or presence of very weak [\ion{O}{ii}]
emission in our spectra (see also N14). 

The spread in the density and temperature values among different objects is the major
source of uncertainty on the derived \mloss . 
In particular, the \oi\, emissivity scales with the density for $n_e \la $10$^6$ \cmt . 
Estimates done on CTT jets base indicate that $n_e$ roughly varies in the range 2-6$\times$10$^{4}$ cm$^{-3}$ and
$T_e$ might be between 7000-15000 K (e.g. Agra-Amboage et al. 2011, 
Maurri et al. 2014 , Hartigan et al. 2007, Giannini et al. 2013). 
Considering the emissivity variations in this range of values, and including a factor of two 
of uncertainty on
the tangential velocity, this leads to a cumulative uncertainty on \mloss\, of about a factor of ten. We remark that electron density and temperature values measured in different jets
do not show any dependence on mass or mass accretion rate of the central stars,
being similar in actively accreting solar mass objects, as well as in brown-dwarfs
and very low mass stars (e.g. Whelan et al. 2009, Ellerbroek et al. 2013). 
This circumstance reassures us that we do not
introduce a bias by assuming the same physical parameters for all sources
of our sample. 

In deriving the mass flux rate, we have summed up both the red-shifted and blue-shifted 
components when detected.
In most of the cases the red-shifted component is not detected, likely because the receding flow is occulted by the disc. To take into account this observational bias,
we have multiplied by two our derived \mloss\, in all the cases where only the
blue-shifted component is detected.
An upper limit on the double-sided \mloss\, have been also estimated from the 
upper limits on the \lhvc\, multiplied by two to correct for the flow bipolarity.

\begin{figure*}
\includegraphics[angle=0,width=15cm, trim = 0.5cm 0 0 4cm,clip]{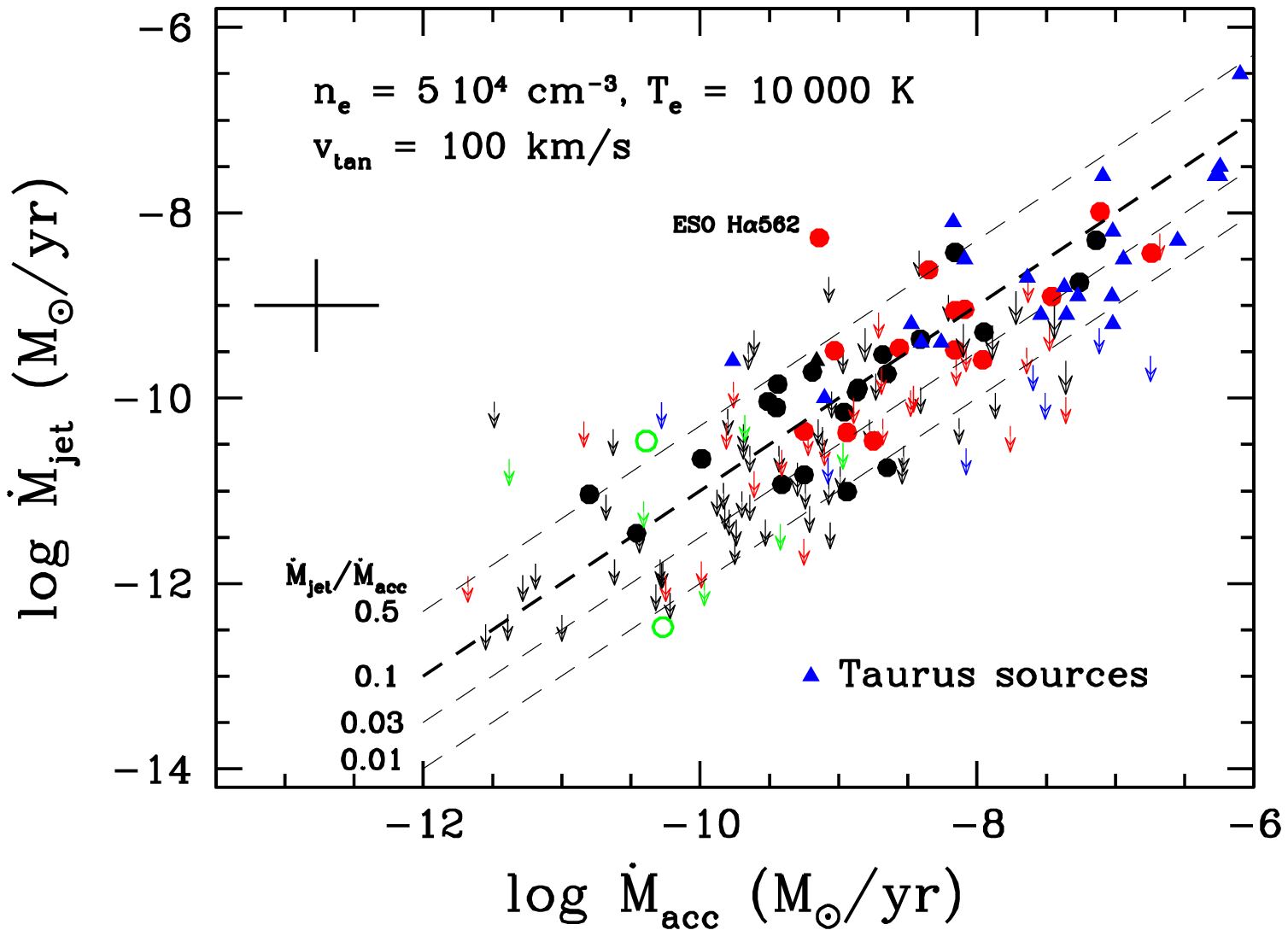}
     \caption{Mass accretion (\.{M}$_{acc}$) vs. mass ejection (\.{M}$_{jet}$) rates. 
     The mass ejection has been measured from the \osei\, luminosity assuming
     a jet tangential velocity of 100\kms\, and gas temperature and density of 10\,000 K and
     5$\times$10$^4$ \cmt\, respectively. Black, red and green symbols
     are as in Fig. \ref{1}, while blue triangles indicate the values for 
     the sample of Taurus sources of H95. Updated mass accretion rates from
     these sources are estimated from the data in Rigliaco et al. 2013. Average data point 
     uncertainty is indicated with a cross on the left side of the plot. Arrows refer to 
     upper limits on \mloss\, derived from the 3$\sigma$ upper limits on the \lhvc\ luminosity.
     }
         \label{maccmloss}
\end{figure*} 

In Figure \ref{maccmloss} the derived mass flux rates are plotted as a function of the 
source mass accretion rates. To guide the eye, lines corresponding to \mloss/\macc\, ratio between 0.01-0.5 are drawn in the figure.
 
In Figure \ref{maccmloss} we also plot the data points relative to the sample
of sources in Taurus whose \mloss\, has been measured by H95
with a similar procedure as the one we adopt here. In H95 the authors
measured the accretion rate from the source veiling, but these values were lately found
to overestimate the true values (Gullbring et al. 1998). Several determinations of the
mass accretion rate of the H95 sample are available in the literature.
For the plot, we have used the determination of the accretion luminosity given by Rigliaco et al (2013)
which derived it from the H$\alpha$ luminosity measured on the same data-set of H95. 
We have then converted \lacc\, into \macc\, by adopting the stellar 
masses and radii given in Rigliaco et al.(2013). The Taurus sources, which extend to higher mass accretion rates with respect to our sample, fall along the same trend as the rest of targets.

The Figure shows that for most of the detections the  \mloss/\macc\, ratio is
between 0.01 and 0.5, confirming, on a large statistical bases, previous results 
found on individual objects. Only one object, ESO-H$\alpha$ 562, has a \mloss/\macc\, ratio much larger than one. This source is a close binary with two components of the same spectral type
but with different extinction values (A$_V$=4 and 10 mag, Daemgen et al. 2013). 
It is therefore possible that the mass accretion rate, estimated from the UV excess, 
refers only to the less extincted of the two stars while the mass ejection rate is the sum of the two
contributions (see discussion in Appendix 3.1 of Manara et al. 2016a). 

\begin{figure}
\includegraphics[angle=0,width=11.7cm, trim = 1.2cm 0 0 7cm,clip]{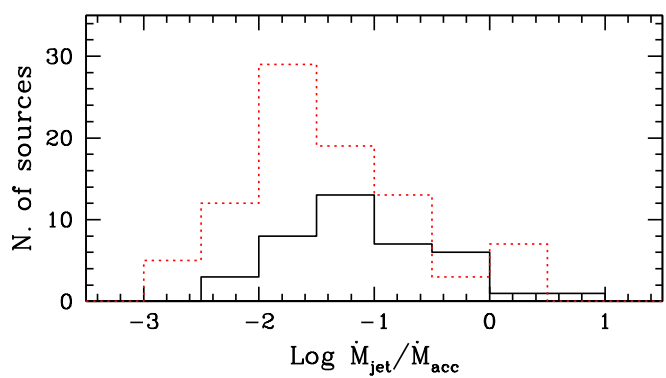}
     \caption{Distribution of ejection efficiency (\mloss/\macc) for the global
     sample. Black histogram are values for sources with detected HVC while
     red dotted histogram indicates the distribution of upper limits.}
         \label{histo_mjet}
\end{figure}

We note a significant number of upper limits pointing
to a very low efficiency of the jet mass loss rate, even for objects with high 
accretion rates. This is better visualized in Fig. \ref{histo_mjet}, which shows an histogram
with the distribution of the estimated  \mloss/\macc\, ratio, dividing detections and
upper limits. Here the Taurus sources are not included.
The peak in the detections is at \mloss/\macc\, = 0.07, with 13/39 (i.e. 33\%) of 
sources in the range 0.03 and 0.1. 
Considering both detections and upper limits, 
however, we find that 57 sources, (44\% of the entire sample), have 
\mloss/\macc\, $<$ 0.03.

\begin{figure}
\includegraphics[angle=0,width=11.7cm, trim = 0cm 0 0 1cm,clip]{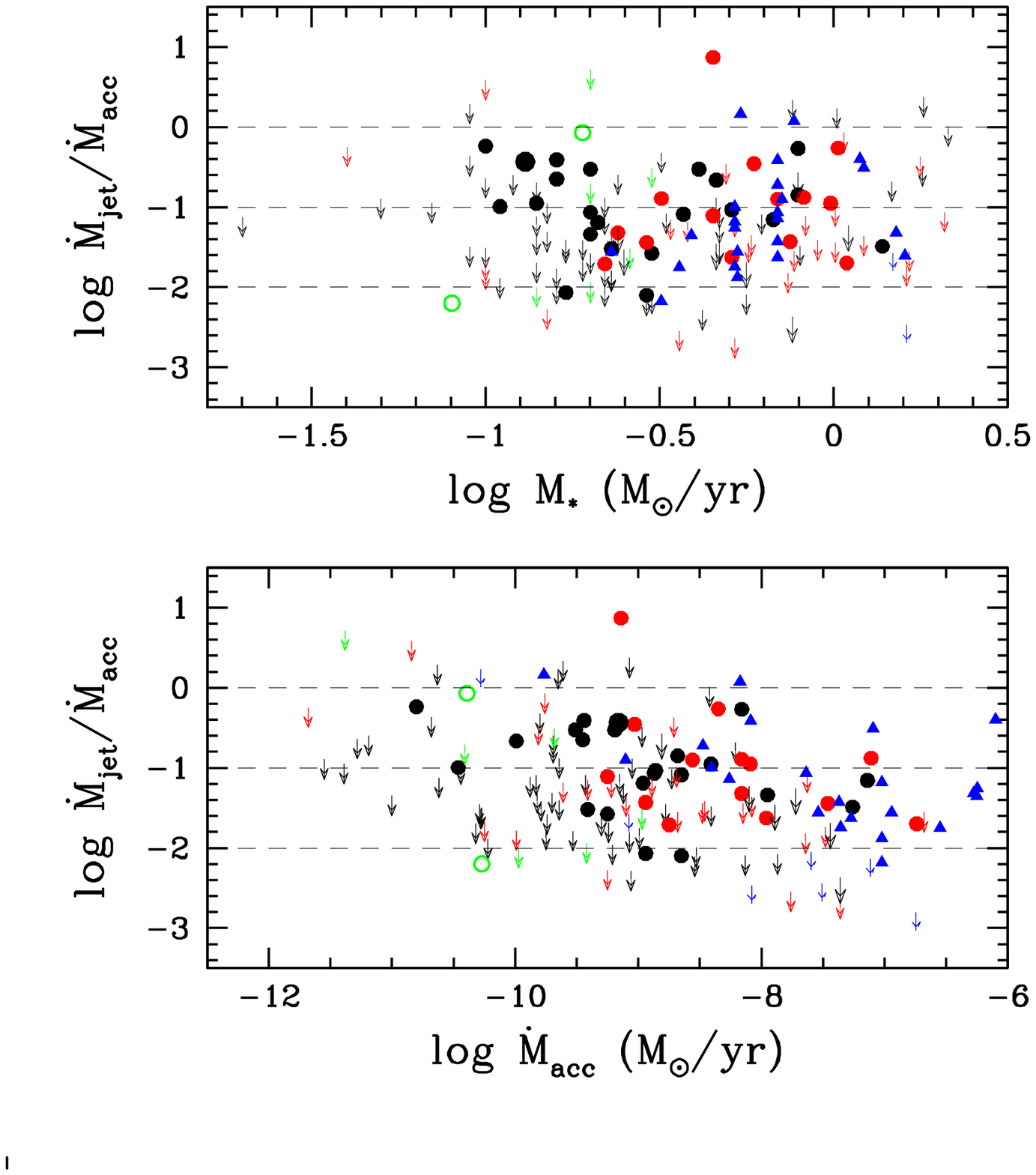}
     \caption{\textbf{Upper panel:} log of the \mloss/\macc\, ratio plotted as a 
     function of the log of the stellar mass. The three horizontal dashed 
     lines correspond to \mloss/\macc\ = 1, 0.1 and 0.01. Black, red and green symbols
     are as in Fig. \ref{1} while blue triangles refer to the sources in Taurus (see caption
     of Fig. \ref{maccmloss}). \textbf{Lower panel}: log of the \mloss/\macc\, ratio plotted as a 
     function of the log of the mass accretion rates. }
         \label{eff}
\end{figure}   

The large scatter in the \mloss/\macc\, ratio is partially due to the uncertainty 
in the determination of the \mloss , and in particular on the assumed parameters. 
However, given the remarkable similarity in the excitation conditions of jets 
from sources with different masses and mass accretion rates, it is likely that the scatter,
of almost two order of magnitude, reflects a real difference in the \mloss\, efficiency 
among sources more than excitation conditions very different from what assumed here.
We also remark that the \mloss/\macc\, ratios for the two TDs of our sample 
with detected HVC are 0.1 and 0.4, that is, in line with the other sources.

The dependence of the \mloss/\macc\, ratio with the mass
and mass accretion rate is presented in Fig. \ref{eff}.  Sources with \macc\, 
between 10$^{-10}$ and 10$^{-8}$\msunyr\, have \mloss/\macc\, ratio that spreads
all values between 0.01 and 1. Lower accretors with detected HVC have a ratio
larger than 0.1, however the non detection of sources with lower values
is likely due to sensitivity limits. Sources with higher mass accretion rates
(i.e. $>$ 10$^{-8}$\msunyr) have a tendency of displaying \mloss/\macc\, ratio
lower than 0.1. The sources in Taurus follow a similar trend
with a few exceptions. 

The upper panel of Fig. \ref{eff} shows that no dependence of the inferred ratio with \mstar\, is observed. Previous studies suggested that the \mloss/\macc\, ratio may be 
higher in BDs than CTTSs (Whelan et al. 2009). In our sample most of sources with 
\mstar\ $\la$ 0.1 \msun\ have upper limits that do not allow us to infer any 
particular trend for the \mloss/\macc\, ratio.

In conclusion, the analysis of \mloss/\macc\, ratio shows that there
is a large spread of values for this ratio that does not depend on the mass of 
the driving source. There is however a tentative trend for sources with 
accretion rates $>$ 10$^{-8}$\msunyr\, to have on average a lower ratio.
This trend needs to be confirmed/rejected on the bases of observations on a 
more complete sample of sources with high mass accretion rates. 

\section{Peak velocities and dispersion}

Our spectral resolution is too low to perform a comprehensive kinematical
analysis of the observed lines. It is however sufficiently high for studying general
trends related to the properties of the gas giving rise to the two velocity components.

\begin{figure}
\includegraphics[angle=0,width=15cm,trim = 1cm 1cm 0 3cm,clip]{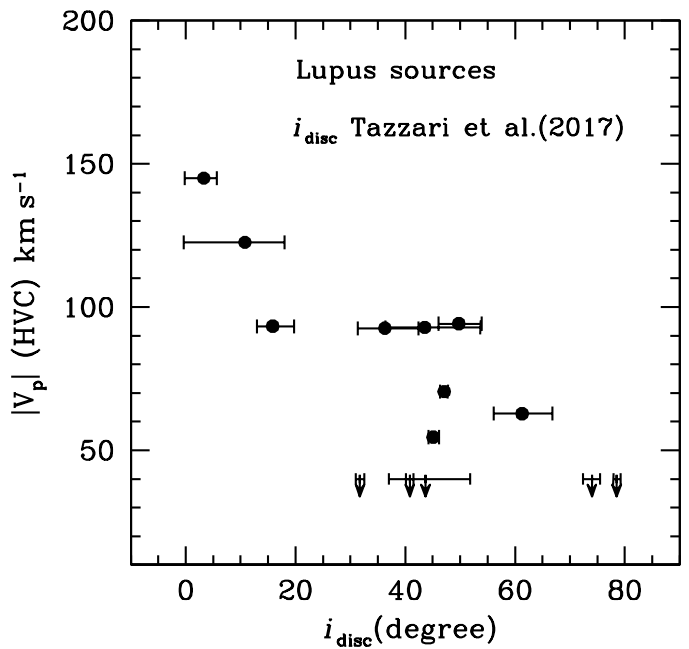}
     \caption{Absolute value of the peak velocity in the HVC (V$_{peak}$(HVC)) is plotted as a function of the
     disc inclination angle for a sub-sample of sources in Lupus where $i_{disc}$
     has been measured from ALMA observations (Tazzari et al. 2017). Upper limits
     refer to sources with measured inclination angles that don't have a HVC component. 
     An upper limit of V$_{peak}$(HVC) of 40\kms\, has been considered in the plot for these
     sources.}
         \label{disc}
\end{figure}

An analysis of the dependence of the observed kinematical behaviour on the jet inclination with respect to the observer
can be done for the sub-sample of the Lupus sources where a measure of the disc inclination
angle is available. Ansdell et al. (2016) observed with ALMA the entire Lupus
sample investigated here, and were able to spatially resolve the disc in 20 sources.
The \osei\, line was detected in 13 of these resolved structures, nine of which 
show HVC emission. Tazzari et al. (2017) revisited the disc inclination angles provided by
Ansdell et al. (2016) applying a more accurate fitting procedure of the visibilities. Here we used these
determinations for our analysis. Fig. \ref{disc} shows the HVC peak velocity (V$_{peak}$(HVC)) plotted as a function of the 
disc inclination angle. The absolute value of the HVC radial projected velocity 
is anti-correlated with the disc inclination angle, as expected in the working
hypothesis that the HVC traces collimated jets ejected in a direction perpendicular
to the disc. In five of the sources where the inclination angle is available we
do not detect a HVC. Two of them have high $i_{disc}$ values (namely Sz84: $i_{disc}$ = 72.9$^o$,
Sz133: $i_{disc}$ = 78.5$^o$) and their position in the plot is compatible with the 
presence of a jet at low radial velocity. Of these sources,  Sz84 has a small LVC width 
($\Delta$V = 23\kms) indicating that if any jet at low radial velocity is present in this source,
its contribution to the overall line profile should be negligible. Sz133, on the other
hand, has a wide \osei\, line ($\Delta$V = 63.7\kms) which is compatible with 
a blended HVC at low radial velocity (see the relative line profile in Fig. 1). 

If we de-project the peak velocities adopting the corresponding inclination angle,
we find total velocities ranging between $\sim$ 100 and 150\kms . Since we consider the
peak of the \osei\, emission, these values represent the mean flow velocity. 
Appenzeller \& Bertout (2013) found a similar correlation in a sample of T Tauri
with known inclination angle, but considering the \osei\, maximum velocity
computed at 25\% of the line intensity. They found for their sample an average jet maximum 
velocity $\sim$ 250\kms . They also performed the same analysis considering the \textit{peak} velocity 
of the [$\ion{N}{ii}$] lines finding jet mean velocities of the order of 200\kms , thus higher
than typical values found by us. This could be related to the fact that the Appenzeller \& Bertout (2013) sample is dominated by more massive sources in the Taurus cloud. Indeed, 
in magneto-centrifugally launched jets, the maximum poloidal velocity is 
roughly proportional to $(r_A/r_0)v_K$, where $r_0$ is
the jet anchoring radius, $r_A$ is the Alfv\'en radius (i.e. the radius
at which the flow poloidal velocity is equal to the Alfv\'en speed), and 
$v_K = \sqrt{(GM_*/r_0)}$ is the Keplerian velocity in the disc at $r_0$.
The lower total jet velocities in the Lupus sources with respect to sources 
in Taurus could be therefore due to their average smaller stellar mass 
and consequently smaller Keplerian velocity and Alfv\'en radius (which in turn
depends on the mass and stellar magnetic field).
Our derived jet velocities are in fact more in line with the velocities found from proper motion 
analysis in a sample of jets of the Cha II cloud (Caratti o Garatti et al. 2009),
driven by sources more similar to those investigated here in terms of stellar parameters .
    
\begin{figure}
\includegraphics[angle=0,width=14cm,trim = 1cm 0 0 6cm,clip]{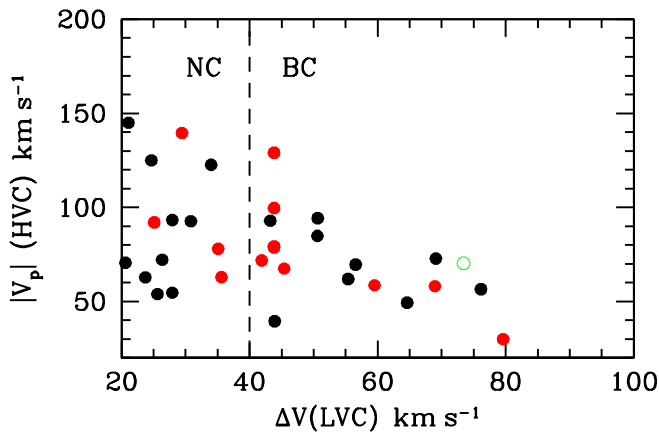}
     \caption{Correlation between V$_{peak}$(HVC) and the deconvolved line width of the LVC. 
     The vertical dashed line at $\Delta$V(LVC) = 40\kms\, separate sources where 
     the LVC is dominated by a narrow component (NC) or a broad component (BC) according
     to Simon et al. (2016). Symbols are as in Fig. 4}
         \label{dv}
\end{figure}

We have also investigated whether there is any dependence among the HVC and LVC
kinematics with the aim of disentagling the origin of the LVC component.
To this aim, Fig. \ref{dv} shows the correlation between V$_{peak}$(HVC) and 
the line width ($\Delta$V) of the LVC. 
The line width has been obtained from the full width at half maximum (FWHM)
 of the fitted Gaussian, deconvolved for the instrumental 
profile ($\Delta V = \sqrt{\Delta V_{obs}^2 - \Delta V_{ins}^2}$).
Sources where the observed line width is smaller than the instrumental resolution are excluded. 
If the LVC emission is dominated by gravitationally bound gas in the disc, we should 
observe an anti-correlation between V$_{peak}$(HVC) and $\Delta$V(LVC), since
a progressively larger spread in disc radial velocities is seen as the disc 
inclination increases. 

High spectral resolution observations have actually shown that the \osei\, LVC 
can be deconvolved into a narrow component (NC) and a broad component (BC), both
roughly centrally peaked (Rigliaco et al. 2013, Simon et al. 2016). Simon et al. (2016), in particular,
found that the LVC-NC has a width roughly $<$ 40\kms\, while the BC have widths 
between 50 and $\sim$ 100\kms . We do not have enough resolution to separate these two
components in our profiles, but we can tentatively assume that sources with LVC FWHM less than 40\kms\, 
are those dominated by the NC while in sources with higher FWHM the BC is progressively more important.
Our plot shows that there is no correlation between V$_{peak}$(HVC) and the LVC width for 
LVC widths less than 40\kms\, while there is a tendency of the V$_{peak}$(HVC) to diminish 
as the LVC width increases in the BC regime. This kind of relationship is expected if the LVC
for these sources have a significant contribution from gas bound in the disc: in this situation,
when the disc is progressively closer to edge-on V$_{peak}$(HVC) decreases while 
the $\Delta$V(LVC) increases.

\section{Discussion}

Our survey of \osei\, emission in a sample of more than 100 T Tauri stars confirms
previous findings that the LVC is a common feature during pre-main sequence evolution, 
being present in sources with a range of accretion luminosity
spanning more than five orders of magnitudes. For most of the sources 
without a detected LVC, the relative upper limits are consistent with
the relationship between \llvc\, and \lacc\, followed by the detected sources.

The frequency of HVC detection is lower, as only 30\% 
of the observed sources show a clear sign of HVC emission (32\% excluding objects with M$_* \la 0.1$). This 
is partially due to the fact that this component is usually weaker with respect to the LVC. 
Although for many of the sources the derived upper limits on the HVC luminosity
are still compatible with the scatter found on the detections, still more
than 50\% of the high accretors (log (\lacc/\lsun) $ > -3$)) do not have 
a clear evidence of high velocity gas.

We should then consider if our choice of \osei\, line as a jet tracer 
introduces some bias in the results. 
This was in part already discussed in H95, who have shown
that the \osei\, luminosity is a good tracer of the total jet mass providing that 
ionization fraction is low, as it is always found in stellar jets.
This is confirmed by the fact that, whenever detected, the \oii\,3726\AA\, 
line is always at least a factor of ten lower than \osei\, implying $x_e <$0.1
(N14). 

The \osei\, line is mainly excited in shocks, therefore one should consider the
possibility that the jet is not radiative because the conditions are not
favourable for shock development in the region probed by our aperture.
Shock excitation occurs due to jet impacts on quiescent ambient medium or 
jet velocity variability that produces an internal clumpy structure.
Shock velocities $\ga$ 20\kms\, are strong enough to sufficiently excite the \osei\, line
at a detectable level (Hartigan et al. 1994).
Shocks occur also in the region of jet recollimation, as 
predicted by MHD models (e.g. Ouyed \& Pudritz 1993) and commonly observed in \osei\, emission 
(e.g. White et al. 2014, Nisini et al. 2016).
Such shocks are stationary and located within $\sim$ 50 AU from the 
source, that is, within 0.2 arcsec at 200 pc, thus detectable in our observational setting. 
In addition to shocks, also ambipolar diffusion in MHD driven jets is efficient enough to heat 
the gas up to $T_e \sim 10\,000$ K still maintaining a low level of ionization (Garcia et al. 2001). 

All the above considerations make it extremely unlikely that jets formed in T Tauri stars 
are not radiative or do not emit in the \osei\, line. 
Consequently, we conclude that sources where stringent upper limits on the \osei\, HVC
have been derived do not develop a jet or the jet 
is weaker than for other sources with similar accretion luminosity. 
We will further discuss this point in Section 8.2.

\subsection{Connection between LVC and HVC}

Both the LVC and HVC correlate better with accretion luminosity than with the 
stellar luminosity or stellar mass. In addition, a correlation between line
and accretion luminosity persists if the two quantities are normalized
to the stellar luminosity. The above finding suggests accretion as the main
driver for line excitation and that the correlation between line
and accretion luminosity is not simply a reflection of the \lacc\, vs. \lstar
correlation, as suggested by Mendigutia et al. (2015).

We also find that the LVC and HVC show a very similar correlation with \lacc\,
and that the \lhvc/\llvc\, ratio does not show any dependence with \lacc .
In the context of a disc-wind origin for the LVC,
this tight connection between the two components is difficult to be explained if the LVC 
mainly originates from a photo-evaporative wind.
In models of photo-evaporative winds the correlation between
\llvc\, and \lacc\, is explained by the underlined correlation between line excitation 
and the EUV flux reaching the disc, this latter being dominated
by accretion photons rather than by stellar photons (Ercolano \& Owen 2016).
On the other hand, the physical mechanism at the origin of the \lhvc\, vs. \lacc\,
correlation is different, being directly related to the \mloss/\macc\, relationship
predicted in any accretion-driven jet-formation model.
Thus in principle the two components should not follow the same relationship
with \lacc\, as they arise from physically distinct mechanisms.
In addition, photo-evaporative winds originate at a larger distance 
in the disc with respect to the jet formation region, therefore 
their properties can be influenced by the jet presence. 
From one side, the jet will likely intercept EUV photons from the 
star-disc interacting region, acting as a veil that prevents 
stellar and/or accretion high energetic photons
to reach the disc surface. Conversely, shocks developing at the jet base
would produce additional energetic photons modifying the 
wind excitation. Therefore, given the different dependence of photo-evaporative winds and
jets on accretion luminosity and the expected feedback effects, it
seems a quite fortuitous coincidence that the two components follow
exactly the same relationship with the accretion luminosity.

Magnetically driven disc-winds models also predict the presence of
a low velocity component formed in the very dense regions at the wind base
(Garcia et al. 2001, Ferreira et al. 2006, Bai et al. 2016, Romanova et al. 2009). 
The interesting aspect of some of these models is that they naturally predict 
the simultaneous presence of a well collimated high velocity jet (the HVC)
which is magneto-centrifugally accelerated in the innermost part of the disc,  
and of a un-collimated wind originating from the outer streamlines (the LVC). 
Line profile predictions for these kind of models do resemble the double component
profile observed in some sources, although the observational details are still poorly matched
(e.g. Shang et al. 1998, Garcia et al. 2001, Pyo et al. 2006).
The possibility that both the HVC and LVC originate from the same physical
mechanism would better explain the tight correlation among the \osei\, 
luminosity of the two components.

\subsection{The \mloss/\macc\, ratio}

The mass loss rate in the jet and the \mloss/\macc\, ratio, are fundamental parameters for the
understanding of the jet formation process and its role in regulating the 
mass transfer from the disc to the young star. In addition, the mass ejected by the jet 
competes with mass loss from low velocity winds 
in dissipating the disc mass, thus regulating the disc evolution.

Several determinations of the \mloss/\macc\, ratio do exist for 
individual sources driving known and bright jets. The derived 
values scatter over a large range, from 0.01 to 1 (see e.g. Ellerbroek et al. 2013
for a collection of values). Often these measurements suffer from
the different methods and assumptions employed for the determination of the accretion 
and ejection properties, or from non-simultaneous observations.
Our statistical approach, based on applying the same methodology on 
a large sample is better suited to give a snapshot of the general property 
of the mass loss over mass accretion ratio for an unbiased population of
T Tauri stars. 
We find that the \mloss/\macc\, ratio ranges from $\la$ 0.01 to $\sim$0.7,
with an average value (from detection only) of $\sim$0.07. 
In addition, and considering the upper limits on the HVC, we show that more than 40\%
of sources have a ratio below 0.03. 
Given the homogeneity of our \mloss\, and \macc\, determinations,
we believe that the observed spread is real and not caused by the 
uncertainty involved on different methodologies to calculate stellar parameters
and mass loss/accretion rates. 
This large range of values, and in particular the very low values and upper limits of \mloss\ found
in several sources, can set interesting constraints on the
existing theoretical models.

Different models have been proposed for the launching and acceleration of
collimated jets and those having a better validation from observations
are the ones involving a magneto-centrifugal acceleration of matter from
the inner disc regions. The core of the mechanism is that matter follows the open magnetic field 
lines threading the disc, and are centrifugally accelerated at super-Alfv\'enic 
speeds. The magnetic coupling between the disc and the wind then transfers
angular momentum outward, allowing disc matter to accrete (see e.g. Blanford \& Payne 1982, 
Hartmann et al. 2016).
Suggested models mainly differ by the region of the star-disc structure from
which the jet originates. This region extends up to few AUs in case of disc-winds
(Ferreira et al. 2006), it is confined to the innermost disc region ($<$ 1 AU) in X-winds
(Shang et al. 2002),
or involves the interaction region between the disc and the stellar magnetosphere
(Zanni \& Ferreira 2013, Romanova et al. 2009, Matt \& Pudritz 2008).
All these models  predict that the \mloss/\macc\, ratio can in principle take
 a wide range of values, from $\sim 10^{-5}$ (Wardle \& K\"onigl 1993) 
 to $\sim$0.6 (Ferreira et al. 2006) thus our observations 
 alone are not sufficient to discriminate between the different scenarios. 
 
In particular, in magneto-centrifugal launching models the \mloss/\macc\, ratio
is roughly proportional to ln($r_{out}/r_{in}$)/$\lambda$,
where $r_{out}$ and $r_{in}$ are the outer and inner radius of the jet launch zone,
and $\lambda = (r_A/r_0)^2$ is the magnetic lever arm (e.g. Hartmann et al. 2016). 
Therefore, the expected values can be inferred by independent observational 
constraints. 
For example, the jet launching region has been constrained in few CTT sources 
through measurements of rotational signatures in optical/UV lines,
 that provide an estimate of the angular momentum
extracted by the jet (e.g. Coffey et al. 2007).
Maximum values in the range $\sim$ 0.6-2 AU have been derived, which imply, assuming 
$r_{in}$ of the order of 0.1 AU, that very low values ($<0.03$) of \mloss/\macc\, 
are attained if $\lambda \ga$ 60 . Such large values of the lever arm are however
not consistent with the relatively low jet terminal velocities estimated in Section 7,
that, for a star of 0.3\msun , would require $\lambda \sim$ 15. 
Therefore, to reconcile our observed large spread of \mloss/\macc\, to the same
disc-wind ejection mechanism, the disc region involved in the
jet acceleration should significantly vary from source to source, and in particular
very small launching zones are required to explain the lowest values of
the observed \mloss/\macc\, ratio.

We should also consider the possibility that 
other effects might concur to reduce the efficiency of the jet acceleration mechanism, at least
in those sources where upper limits point to an extremely low \mloss/\macc\, ratio.
Ultimately, the conditions for having an efficient launching mechanism
are mainly settled by the actual morphology of the local magnetic field responsible 
for the magneto-centrifugal acceleration.
Models usually assume an axi-symmetric dipolar field, while the real field 
topology could be very different. Observations based on 
Zeeman-Doppler imaging studies have in fact shown that the stellar magnetic field in T Tauri
stars can present a variety of large-scale geometries, from simple and axi-symmetric 
to complex and non-axisymmetric (e.g. Donati et al. 2007, Hussain et al. 2009). 
Gregory et al. (2012, 2016) have suggested that the magnetic field topology 
might depend on the stellar internal structure, a result based on the
finding that in PMS stars more massive than 0.5\msun\, 
the ratio between the strength of the octupole
to the dipole component of the magnetic field increases with age, as the star
passes from being fully convective into developing a radiative core. 
They suggest that even less massive stars, that are fully convective during 
all their PMS life, could present a variety of magnetic field topologies,
on the basis of their similarity with the magnetic properties of main-sequence
M-dwarfs. How the stellar magnetic field configuration influences the properties 
of the jet launching region in the inner disc is still not properly addressed. 
Ferreira (1997) for example showed that in MHD disc-wind models a dominant 
quadrupolar magnetic field configuration leads to a much weaker wind than a 
dipolar configuration. On the other hand, Mohanty \& Shu (2008) modelled the 
outflow launch in the presence of a multipole magnetic field under
the framework of X-wind theory, finding that it is little affected by the 
field configuration.  This is certainly a subject that deserves more investigations.

The onset of a collimated jet could be also influenced by the orientation of the 
disc with respect to the local interstellar magnetic field. Menard \& Duchene (2004)
observed that CTT discs are randomly oriented with respect to the ambient magnetic field,
but suggested that disc sources without any bright and extended
outflow have a tendency to align perpendicularly to the magnetic field.
This in turn may indicate that these sources have a less favourable 
topology for the magnetic field in the inner disc, resulting from the interaction
between the ambient and the stellar fields.
Targon et al. (2011) performed a similar analysis studying the alignment 
between the interstellar magnetic field and the jet direction for
sources at different evolutionary stages. They found that
jets from classes 0 and I align better than T Tauri stars. 

In this framework, it would be interesting to investigate for variations of
the \mloss/\macc\, ratio with age to understand if
the jet ejection efficiency depends on protostellar evolution.
Unfortunately, the statistics on the mass ejection/mass accretion ratio in younger
sources (e.g. class0/I sources) is very limited mainly due to the
intrinsic difficulty of estimating the mass accretion rate in the
embedded phase. Observations performed on few well known class I jet driving sources
(Antoniucci et al. 2008, Nisini et al. 2016) suggest a large variety of \mloss/\macc\, ratios
spanning between 0.01 and 0.9, similarly to what we measure here. Large, unbiased 
statistics on the real occurrence of jets in class I phase are lacking, 
through White et al. (2004) observed optical fobidden line emission through 
scattered light in 15 class I sources in Taurus-Auriga, finding that only five show evidence 
of a HVC, which is a detecting rate similar to that found 
in our sample.  On the other hand, it is rather well established that all known class 0 sources 
have mass ejection in the form of bipolar outflows indirectly pointing to a collimated 
jet as the driving agent. Nisini et al. (2015) directly resolved 
the atomic jet in a small sample of five class 0 sources through 
\textit{Herschel} observations of the \oi\,63\um\, line. Assuming that 
the source accretion luminosity is comparable to the bolometric luminosity they 
derived a \mloss/\macc\, efficiency in the range 0.05-0.5.
These efficiencies can be actually higher if a significant fraction
 of the mass in the jet is in molecular form, as many sub-mm observations suggest, and
 if accretion luminosity accounts for only a fraction of the source bolometric luminosity. 
 Although on a very limited sample, these findings suggest that jets are more
 efficient in transferring mass and momentum  outwards in early stages 
 of protostellar evolution while the effiency diminishes with time. 

\section{Conclusions}

We have presented a study of the \osei\, line in a sample of 131 young stars
with discs in the Lupus, Chamaeleon and $\sigma$ Orionis star forming regions,
having mass accretion rates spanning from 10$^{-12}$ to 10$^{-7}$ \msunyr . 
The line has been deconvolved into the two kinematical components, that is the LVC peaking
close to zero velocity and the HVC, associated with high velocity jets,
 with velocity shifts $>$40\kms. The LVC is detected in 77\% of the sources, while
 the HVC is present in only 31\% of the objects. We have correlated the luminosity 
 of the two line components with different stellar and accretion parameters of the sources.
 In addition, we have estimated, from the HVC luminosity, the mass ejection rate 
 (\mloss) in the jet, deriving the distribution of \mloss/\macc\, jet efficiency ratio in 
 our sample. Velocity shifts and line widths have been
 analysed and compared in order to infer connections between the two components
 related to their origin. The main results from our study are the following:
 
\begin{itemize}

\item After having evaluated geometrical and sensitivity effects, we conclude that
mass losses in the form of jets (i.e. the HVC) are less frequent than slow- winds
(i.e. the LVC) at least in the class II phase investigated here. The HVC is 
on average more often detected in sources showing high accretion luminosity (i.e.
log (\lacc/\lsun) $\ga -$3 , detection rate 39\%).

\item The \osei\, luminosity of both the LVC and HVC, when detected, correlates with 
stellar and accretion parameters of the central sources (i.e. \lstar\, \mstar\, \lacc, \macc),
with similar slopes for the two components. Line luminosity correlates better with
accretion luminosity than with the stellar luminosity or stellar mass. In addition,
a tight correlation is still found in a plot \loi/\lstar\, vs. \lacc/\lstar. 
On these basis we suggest that accretion is the main driver for line excitation.

\item We find an inverse correlation between the peak velocity of the HVC 
and the disc inclination angle measured in the sub-sample of Lupus sources 
observed by ALMA. This confirms our working hypothesis that the HVC traces
collimated jets ejected in a direction perpendicular to the disc plane. 
The inferred average jet velocity, corrected for inclination angles, is between
100 and 150\kms\, that is about a factor of two lower than 
typical jet velocities estimated in Taurus sources. We suggest this is due to
a dependence of the total jet velocity on the mass of the central source. 

\item The very similar correlations of \llvc\, and \lhvc\, and
the accretion luminosity suggests a common mechanism for the formation of
the LVC and HVC. This supports the idea that magnetically driven disc-winds
are at the origin of both components, as they can predict the simultaneous
presence of collimated high velocity jets, accelerated in the innermost
part of the disc, and un-collimated slow winds, originating from the
outer streamlines. A contribution to the LVC from gas still bound to the disc  
is also suggested in objects with large $\Delta$V(LVC).

\item Mass ejection rates (\mloss) measured from the \lhvc\, span from $\sim$ 10$^{-13}$
to $\sim$ 10$^{-7}$ \msunyr\, for the sub-sample with a HVC detection. 
The corresponding \mloss/\macc\, ratio ranges from $\sim$0.01 to 0.5, with an average
value of 0.07. However, considering also upper limits on the HVC, we find that 
$\sim$ 40\% of sources of the total sample has a \mloss/\macc\, ratio $<$ 0.03.
There is a tentative evidence that sources with higher \macc\, have on average a
lower \mloss/\macc\, ratio, that needs to be confirmed on a larger sample
of strong accretors. 
 
\item The observed large spread in the \mloss/\macc\, ratio poses
constraints on the existing 
theoretical models, implying that the disc region from where
the jet is launched significantly varies from source to source.
The very low values and upper limits of the \mloss/\macc\, ratio found
in several sources, in particular, point to very small jet 
launching zones in these sources, if one considers 
standard MHD models. 
An alternative hypothesis is that the sources of our sample
with stringent upper limits in the \mloss/\macc\, ratio do not have
the conditions for the development of a high velocity jet in their star-disc
interaction region. This might be the case if the configuration of the magnetic field 
 in these stars is not optimized for an efficient magneto-centrifugal 
acceleration in the inner disc.  

\end{itemize}

Our survey of \osei\, emission shows that a significant fraction of 
class II discs with typical ages of 2-4 Myr, as those in Lupus and Chamaeleon, 
do not show prominent high velocity jets as one would have expected given their
accretion activity level. At variance, slow winds seem more efficiently produced 
and could potentially represent the main agent for disc mass loss in this evolutionary stage.
Systematic observations of both mass accretion and mass ejection rates in 
populations of younger and embedded stars, (in the class0 and I 
stage), are needed to determine the role of jets during protostellar evolution.
Such surveys might become possible in the close future thanks to the JWST
facility. On a longer term, this kind of study could benefit from 
space instruments like SPICA/SMI, 
working between 12-18 \um\, with a resolution (R$\sim$30\,000) high enough 
to separate the LVC and HVC in mid-IR forbidden lines. 

\begin{acknowledgements}
We acknowledge fruitful discussions with members of the JEDI (JEts \& discs@INAF) group. 
We thank Marco Tazzari for sharing the information of the Lupus disc inclination
angles beforehand publication. This work has been supported by the PRIN INAF 2013 `discs, jets and the dawn of planets'.
AN acknowledges funding from Science Foundation Ireland (Grant 13/ERC/I2907). DF acknowledges support from the Italian Ministry of Education, Universities and Research, project SIR (RBSI14ZRHR).
C.F.M. acknowledges an ESA research fellowship. 
This research made use of the SIMBAD database, operated at CDS, Strasbourg, France.

\end{acknowledgements}


\begin{appendix}

\section{Cha II new sample and parameters}

The X-shooter spectra of a sub-sample of ChaII sources used in this paper have been
analysed here for the first time, thus we briefly describe 
the observations, data reduction and derivation of stellar and accretion
parameters.
The data belong to the sample observed in Pr.Id. 090.C-0253 (PI Antoniucci). Source names
and coordinates are listed in Table \ref{tab_b1}. Targets were acquired with 
the narrow X-shooter slits, providing resolution of $R \sim$ 18\,000 in the VIS arm.
No exposures with the large slit were obtained, thus spectra have been flux calibrated
using available photometric data. In practice we first match the spectra from the three arms 
and then scale all of them to minimize the scatter to the photometric points. 

Stellar and accretion parameters, listed in Table \ref{tab_b1}, have been self-consistently 
derived from the X-shooter spectra adopting the automatic procedure described in Manara et al. (2013),
which has been also used for all the other sources of this
survey, as briefly discussed in Section 2. 
Masses are estimated adopting the tracks by Baraffe et a. (2015).

\begin{footnotesize}

\begin{table*}[h]
\label{tab_b1}
\caption[]{Name, coordinates and derived properties for the new Chamaeleon II sources}
\vspace{0.5cm}
\begin{tabular}{lcccccccccc}
\hline
Source & Other names & RA(2000)  & DEC(2000) & SpT & $T_{eff}$ & A$_V$ & $L_{\star}$ & log$L_{acc}$ &$M_\star$ & log$\dot M_{\rm acc}$\\
       &             &           &           &     &  K      & mag &  \lsun      &  \lsun
& \msun    & \msunyr \\
\\
\hline
\\
Sz49   & ISO-ChaII 55/CHIIXR 9  & 13:00:53.23& -76:54:15.2 & M3 & 3415 & 0.8     & 0.11 & $-$2.01 & 0.32 & $-$8.94\\
Sz48SW &  &           13:00:53.56 & -77:09:08.3    & M1 & 3705 & 3.7    & 0.27 & $-$2.83 & 0.49   & $-$9.81\\
CMCha  & CHIIXR~13  & 13:02:13.57 & -76:37:57.8  & K7 & 4060 & 0.4 & 0.51 & $-$1.55 & 0.77 & $-$8.68 \\
Hn22   &  &           13:04:22.85 & -76:50:05.5    & M2 & 356   & 0.4   & 0.21 & $-$2.71 & 0.38   & $-$9.61 \\
Hn23   &  &           13:04:24.11 & -76:50:01.2    & K5 & 4350 & 0.8    & 1.11 & $-$1.55 & 1.01   & $-$8.69\\
Sz53   &  &           13:05:12:69& -77:30:52.3  & M3 & 3415 & 2.8       & 0.11 & $-$1.23 & 0.32   & $-$8.16\\
Sz58   &  &           13:06:57.45& -77:23:41.55    & K6 & 4205 & 3.5    & 0.53 & $-$1.24&         0.90& $-$8.46\\ 
Sz61   &  BM~Cha/HBC~594 &  13:08:06.28& -77:55:05.2    & K5 & 4350 & 3.1       & 1.61 & $-$1.05 & 0.98   & $-$8.09\\ 
\\
\hline\\[-5pt]
\end{tabular}
\end{table*} 
\end{footnotesize}

\section{Tables of source parameters and line fluxes.}

\newpage
\clearpage
\onecolumn
\begin{longtable}{lccccccccc}
\caption[]{Parameters of the objects}\\
\hline
Source & RA & DEC & $d$ & A$_V$ & $L_\star$ & log($L_{acc}$) & $M_\star$ & log($M_{acc}$) &\\
       &    &      &  pc   &  mag   &  L$_\odot$ & L$_\odot$   & \msun  & \msunyr & Notes\\
\hline\\[+5pt]  
\endfirsthead
\caption{Continued.} \\
\hline
Source & RA & DEC & $d$ & A$_V$ & $L_\star$ & log($L_{acc}$) & $M_\star$ & log($M_{acc}$) &\\
       &    &      &  pc   &  mag   &  L$_\odot$ & L$_\odot$   & \msun  & \msunyr & Notes \\
\hline\\[+5pt]  
\endhead
\multicolumn{10}{c}{Lupus GTO sample}\\
\hline\\[+5pt]
 Sz~66             & 15:39:28.28   & $-$34:46:18.0    &  150    &  1.0    &   0.200  &   $-$1.8   &   0.29    &    $-$8.54 &   \\ 
 AKC2006-I         & 15:44:57.90   & $-$34:23:39.5    &  150    &  0.0    &   0.016  &   $-$4.1   &   0.14    &    $-$11.0 &   \\
 Sz~69             & 15:45:17.42   & $-$34:18:28.5    &  150    &  0.0    &   0.088  &   $-$2.8   &   0.20    &    $-$9.51 &  \\
 Sz~71             & 15:46:44.73   & $-$34:30:35.5    &  150    &  0.5    &   0.309  &   $-$2.2   &   0.42    &    $-$9.06 &  \\
 Sz~72             & 15:47:50.63   & $-$35:28:35.4    &  150    &  0.7    &   0.252  &   $-$1.8   &   0.37    &    $-$8.65 &   \\
 Sz~73             & 15:47:56.94   & $-$35:14:34.8    &  150    &  3.5    &   0.419  &   $-$1.0   &   0.79    &    $-$8.16 &   \\
 Sz~74             & 15:48:05.23   & $-$35:15:52.8    &  150    &  1.5    &   1.043  &   $-$1.5   &   0.30    &    $-$7.87 &  \\
 Sz~83             & 15:56:42.31   & $-$37:49:15.5    &  150    &  0.0    &   1.313  &   $-$0.3   &   0.67    &    $-$7.14 &  \\
 Sz~84             & 15:58:02.53   & $-$37:36:02.7    &  150    &  0.0    &   0.122  &   $-$2.7   &   0.16    &    $-$9.21 & TD \\
 Sz~130            & 16:00:31.05   & $-$41:43:37.2    &  150    &  0.0    &   0.160  &   $-$2.2   &   0.41    &    $-$9.19 &  \\
 Sz~88A            & 16:07:00.54   & $-$39:02:19.3    &  200    &  0.2    &   0.488  &   $-$1.2   &   0.56    &    $-$8.13 &  \\
 Sz~88B            & 16:07:00.62   & $-$39:02:18.1    &  200    &  0.0    &   0.118  &   $-$3.1   &   0.20    &    $-$9.74 &  \\ 
 Sz~91             & 16:07:11.61   & $-$39:03:47.1    &  200    &  1.2    &   0.311  &   $-$1.8   &   0.47    &    $-$8.73 & TD \\ 
 Lup713            & 16:07:37.72   & $-$39:21:38.8    &  200    &  0.0    &   0.020  &   $-$3.5   &   0.11    &    $-$10.22 &  \\
 Lup604s           & 16:08:00.20   & $-$39:02:59.7    &  200    &  0.0    &   0.057  &   $-$3.7   &   0.14    &    $-$10.32 &   \\ 
 Sz~97             & 16:08:21.79   & $-$39:04:21.5    &  200    &  0.0    &   0.169  &   $-$2.9   &   0.23    &    $-$9.53 &   \\ 
 Sz~99             & 16:08:24.04   & $-$39:05:49.4    &  200    &  0.0    &   0.074  &   $-$2.6   &   0.23    &    $-$9.41 &   \\
 Sz~100            & 16:08:25.76   & $-$39:06:01.1    &  200    &  0.0    &   0.169  &   $-$3.0   &   0.16    &    $-$9.44 & TD \\ 
 Sz~103            & 16:08:30.26   & $-$39:06:11.1    &  200    &  0.7    &   0.188  &   $-$2.4   &   0.22    &    $-$8.99 &   \\ 
 Sz~104            & 16:08:30.81   & $-$39:05:48.8    &  200    &  0.0    &   0.102  &   $-$3.2   &   0.16    &    $-$9.75 &   \\ 
 Lup706            & 16:08:37.30   & $-$39:23:10.8    &  200    &  0.0    &   0.003  &   $-$4.8   &   0.05    &    $-$11.55 & SL \\ 
 Sz~106            & 16:08:39.76   & $-$39:06:25.3    &  200    &  1.0    &   0.098  &   $-$2.5   &   0.62    &    $-$9.83 & SL\\ 
 Par-Lup3-3        & 16:08:49.40   & $-$39:05:39.3    &  200    &  2.2    &   0.240  &   $-$2.9   &   0.22    &    $-$9.43 &  \\ 
 Par-Lup3-4        & 16:08:51.43   & $-$39:05:30.4    &  200    &  0.0    &   0.003  &   $-$4.1   &   0.17    &    $-$11.49 & SL\\
 Sz~110            & 16:08:51.57   & $-$39:03:17.7    &  200    &  0.0    &   0.276  &   $-$2.0   &   0.22    &    $-$8.53 &    \\ 
 Sz~111            & 16:08:54.69   & $-$39:37:43.1    &  200    &  0.0    &   0.330  &   $-$2.2   &   0.47    &    $-$9.12 & TD\\ 
 Sz~112            & 16:08:55.52   & $-$39:02:33.9    &  200    &  0.0    &   0.191  &   $-$3.2   &   0.17    &    $-$9.64 & TD\\ 
 Sz~113            & 16:08:57.80   & $-$39:02:22.7    &  200    &  1.0    &   0.064  &   $-$2.1   &   0.20    &    $-$8.87 &  \\ 
 2MASSJ16085953    & 16:08:59.53   & $-$38:56:27.6    &  200    &  0.0    &   0.009  &   $-$4.6   &   0.02    &    $-$10.62   &   \\ 
 SSTc2dJ160901.4   & 16:09:01.40   & $-$39:25:11.9    &  200    &  0.5    &   0.148  &   $-$3.0   &   0.22    &    $-$9.64 &  \\ 
 Sz~114            & 16:09:01.84   & $-$39:05:12.5    &  200    &  0.3    &   0.312  &   $-$2.5   &   0.21    &    $-$8.96 &  \\ 
 Sz~115            & 16:09:06.21   & $-$39:08:51.8    &  200    &  0.5    &   0.175  &   $-$2.7   &   0.19    &    $-$9.24 &  \\
 Lup-818           & 16:09:56.29   & $-$38:59:51.7    &  200    &  0.0    &   0.025  &   $-$4.1   &   0.09    &    $-$10.68 &  \\ 
 Sz~123A           & 16:10:51.34   & $-$38:53:14.6    &  200    &  1.2    &   0.203  &   $-$1.8   &   0.51    &    $-$8.86 & TD\\ 
 Sz~123B           & 16:10:51.31   & $-$38:53:12.8    &  200    &  0.0    &   0.051  &   $-$2.7   &   0.46    &    $-$9.99 & SL \\ 
 SST-Lup3-1        & 16:11:59.81   & $-$38:23:38.5    &  200    &  0.0    &   0.059  &   $-$3.6   &   0.17    &    $-$10.29 &   \\ 
\hline\\
\multicolumn{10}{c}{Lupus New Sample}\\
\hline\\
 Sz65             & 15:39:27.78   & $-$34:46:17.4    &  150    &   0.6   &   0.832  &   $<-$2.6   &   0.70   &   $<-$9.57 & CH \\ 
 AKC2006-18       & 15:41:40.82   & $-$33:45:19.0    &  150    &   0.0   &   0.011  &   $-$4.6   &    0.07   &   $-$11.24  &    \\ 
 SSTc2dJ154508.9  & 15:45:08.88   & $-$34:17:33.7    &  150    &   5.5   &   0.057  &   $-$1.8   &    0.14   &   $-$8.41  &    \\ 
 Sz68             & 15:45:12.87   & $-$34:17:30.8    &  150    &   1.0   &   5.128  &   $<-$1.2   &   2.13   &   $<-$8.42  & CH\\ 
 SSTc2dJ154518.5  & 15:45:18.53   & $-$34:21:24.8    &  150    &   0.0   &   0.041  &   $-$4.3   &    0.08   &   $-$10.70  &    \\
 Sz81A            & 15:55:50.21   & $-$38:01:34.0    &  150    &   0.0   &   0.224  &   $-$2.5   &    0.19   &   $-$8.98  &    \\ 
 Sz81B            & 15:55:50.26   & $-$38:01:32.2    &  150    &   0.0   &   0.110  &   $-$3.2   &    0.14   &   $-$9.70  &    \\ 
 Sz129            & 15:59:16.48   & $-$41:57:10.3    &  150    &   0.9   &   0.371  &   $-$1.2   &    0.79   &   $-$8.32  &    \\ 
 SSTc2dJ155925.2  & 15:59:25.24   & $-$42:35:07.1    &  150    &   0.0   &   0.019  &   $-$4.4   &    0.14   &   $-$11.26  &    \\ 
 RYLup            & 15:59:28.39   & $-$40:21:51.3    &  150    &   0.4   &   1.660  &   $-$0.9   &    1.40   &   $-$8.19 & TD \\ 
 SSTc2dJ160000.6  & 16:00:00.62   & $-$42:21:57.5    &  150    &   0.0   &   0.087  &   $-$3.1   &    0.20   &   $-$9.81  &   \\  
 SSTc2dJ160002.4  & 16:00:02.37   & $-$42:22:15.5    &  150    &   1.4   &   0.148  &   $-$3.0   &    0.22   &   $-$9.66  &    \\ 
 SSTc2dJ160026.1  & 16:00:26.13   & $-$41:53:55.6    &  150    &   0.9   &   0.066  &   $-$3.3   &    0.14   &   $-$9.88  &    \\ 
 MYLup            & 16:00:44.53   & $-$41:55:31.2    &  150    &   1.3   &   0.776  &   $<-$2.3   &   1.06   &   $<-$9.67   &  TD, CH \\ 
 Sz131            & 16:00:49.42   & $-$41:30:04.1    &  150    &   1.3   &   0.138  &   $-$2.4   &    0.32   &   $-$9.28  &    \\ 
 Sz133            & 16:03:29.41   & $-$41:40:02.7    &  150    &   1.8   &   0.071  &   $-$2.4   &    ...    &   $-$...   &   SL   \\ 
 SSTc2dJ160703.9 & 16:07:03.84   & $-$39:11:11.3    &  200    &   0.6   &   0.005  &   $-$5.2   &    0.17   &   $-$12.38 & SL\\
 SSTc2dJ160708.6  & 16:07:08.55   & $-$39:14:07.6    &  200    &   3.6   &   0.011  &   $-$2.2   &    0.13   &   $-$8.64  &    \\ 
 Sz90             & 16:07:10.08   & $-$39:11:03.5    &  200    &   1.8   &   0.661  &   $-$1.6   &    0.73   &   $-$8.64  &    \\  
 Sz95             & 16:07:52.32   & $-$38:58:06.3    &  200    &   0.8   &   0.417  &   $-$2.5   &    0.29   &   $-$9.09  &    \\ 
 Sz96             & 16:08:12.62   & $-$39:08:33.5    &  200    &   0.8   &   0.692  &   $-$2.3   &    0.43   &   $-$9.02  &    \\ 
 2MASSJ16081497   & 16:08:14.96   & $-$38:57:14.5    &  200    &   1.5   &   0.009  &   $-$3.4   &    0.10   &   $-$10.27   &  \\ 
 Sz98             & 16:08:22.50   & $-$39:04:46.0    &  200    &   1.0   &   2.512  &   $-$0.5   &    0.70   &   $-$7.23  &    \\ 
 Lup607           & 16:08:28.10   & $-$39:13:10.0    &  200    &   0.0   &   0.071  &   $<-$4.9   &   0.10   &    $<-$11.28 & CH \\ 
 Sz102            & 16:08:29.73   & $-$39:03:11.0    &  200    &   0.7   &   0.015  &   $-$2.0   &    ...    &    ...  & SL  \\
 SSTc2dJ160830.7  & 16:08:30.70   & $-$38:28:26.8    &  200    &   0.2   &   3.020  &   $<-$1.8   &   1.81   &    $<-$9.96 & TD, CH \\ 
 V1094Sco        & 16:08:36.18   & $-$39:23:02.5    &  200    &   1.7   &   1.95   &   $-$0.8   &    0.79   &    $-$7.67  &    \\ 
 Sz108B           & 16:08:42.87   & $-$39:06:14.7    &  200    &   1.6   &   0.151  &   $-$2.9   &    0.17   &    $-$9.40   &   \\ 
 2MASSJ16085324   & 16:08:53.23   & $-$39:14:40.3    &  200    &   1.9   &   0.302  &   $-$3.1   &    0.29   &    $-$9.76  &    \\ 
 2MASSJ16085373   & 16:08:53.73   & $-$39:14:36.7    &  200    &   4.0   &   0.007  &   $-$3.7   &    0.10   &    $-$10.63  &   \\ 
 2MASSJ16085529   & 16:08:55.29   & $-$38:48:48.1    &  200    &   0.0   &   0.076  &   $-$4.1   &    0.11   &    $-$10.51   &   \\ 
 SSTc2dJ160927.0  & 16:09:26.98   & $-$38:36:27.6    &  200    &   2.2   &   0.115  &   $-$1.3   &    0.19   &    $-$7.93  &    \\ 
 Sz117            & 16:09:44.34   & $-$39:13:30.3    &  200    &   0.5   &   0.447  &   $-$2.1   &    0.26   &    $-$8.61  &    \\ 
 Sz118            & 16:09:48.64   & $-$39:11:16.9    &  200    &   1.9   &   1.071  &   $-$1.8   &    1.01   &    $-$8.94  &    \\ 
 2MASSJ16100133   & 16:10:01.32   & $-$39:06:44.9    &  200    &   1.7   &   0.209  &   $-$3.4   &    0.10   &    $-$9.55  &    \\  
 SSTc2dJ161018.6  & 16:10:18.56   & $-$38:36:13.0    &  200    &   0.5   &   0.060  &   $-$3.8   &    0.17   &    $-$10.50  &    \\
 SSTc2dJ161019.8  & 16:10:19.84   & $-$38:36:06.8    &  200    &   0.0   &   0.071  &   $-$3.9   &    0.10   &    $-$10.28  &    \\ 
 SSTc2dJ161029.6  & 16:10:29.57   & $-$39:22:14.7    &  200    &   0.9   &   0.158  &   $-$3.2   &    0.19   &    $-$9.76  & TD \\ 
 SSTc2dJ161243.8  & 16:12:43.75   & $-$38:15:03.3    &  200    &   0.8   &   0.617  &   $-$2.0   &    0.44   &    $-$8.76  &    \\ 
 SSTc2dJ161344.1  & 16:13:44.11   & $-$37:36:46.4    &  200    &   0.6   &   0.069  &   $-$2.3   &    0.16   &    $-$8.94  &    \\ 
 \hline\\
\multicolumn{10}{c}{Lupus Archive}\\
\hline\\
 Sz75/GQ\,Lup         & 15:49:12.10  & $-$35:39:05.1 &  150   & 0.70     & 1.44     &  $-$0.7 &    0.86  &  $-$7.67    &      \\
 Sz76                 & 15:49:30.74  & $-$35:49:51.4 &  150   & 0.20     & 0.16     &  $-$2.6 &    0.23  &  $-$9.26    &  TD  \\
 Sz77                 & 15:51:46.95  & $-$35:56:44.1 &  150   & 0.00     & 0.55     &  $-$1.7 &    0.75  &  $-$8.79    &      \\
 RXJ1556.1-3655       & 15:56:02.09  & $-$36:55:28.3 &  150   & 1.00     & 0.23     &  $-$0.9 &    0.50  &  $-$7.92    &      \\
 Sz82/IM\,Lup         & 15:56:09.18  & $-$37:56:06.1 &  150   & 0.90     & 2.33     &  $-$1.1 &    0.95  &  $-$8.04    &  TD  \\
 EX\,Lup              & 16:03:05.49  & $-$40:18:25.4 &  200   & 1.10     & 1.23     &  $-$0.7 &    0.52  &  $-$7.41    &      \\
\hline\\
\multicolumn{10}{c}{Chamaeleon I}\\
\hline\\
  Ass~Cha~T~2-3     & 10:55:59.73   & $-$77:24:39.9    &   160  &   2.6  &   0.18    &  $-$1.05   &   0.79  &    $-$8.43 &  \\ 
  TW~Cha            & 10:59:01.09   & $-$77:22:40.7    &   160  &   0.8  &   0.38    &  $-$1.66   &   1.00  &    $-$8.96 &  \\  
  CR~Cha            & 10:59:06.97   & $-$77:01:40.3    &   160  &   1.3  &   3.26    &  $-$1.42   &   1.78  &    $-$8.71 &  \\ 
  Ass~Cha~T~2-12    & 11:02:55.05   & $-$77:21:50.8    &   160  &   0.8  &   0.15    &  $-$2.12   &   0.23  &    $-$8.77 &  \\ 
  CT~Cha~A          & 11:04:09.09   & $-$76:27:19.3    &   160  &   2.4  &   1.50    &  $-$0.30   &   1.40  &    $-$7.02 &  \\ 
  Hn~5              & 11:06:41.81   & $-$76:35:49.0    &   160  &   0.0  &   0.05    &  $-$2.56   &   0.16  &    $-$9.27 &  \\ 
  Ass~Cha~T~2-23    & 11:06:59.10   & $-$77:18:53.6    &   160  &   1.7  &   0.32    &  $-$1.65   &   0.33  &    $-$8.29 &  \\ 
  Ass~Cha~T~2-24    & 11:07:12.07   & $-$76:32:23.2    &   160  &   1.5  &   0.40    &  $-$1.48   &   0.91  &    $-$8.68  & \\ 
  Cha$-$H$\alpha-$1 & 11:07:16.69   & $-$77:35:53.3    &   160  &   0.0  &   0.01    &  $-$5.11   &   0.05  &    $-$11.69 &  \\ 
  Cha$-$H$\alpha-$9 & 11:07:18.61   & $-$77:32:51.7    &   160  &   4.8  &   0.03    &  $-$4.19   &   0.1   &    $-$10.85  & \\  
  Sz~19             & 11 07 20.72   & $-$77 38 07.3    &   160  &   1.5  &    5.1    &  $-$0.37   &   2.08  &    $-$7.63  &  \\ 
  Sz~22             & 11:07:57.93   & $-$77:38:44.9    &   160  &   3.2  &   0.51    &  $-$0.92   &   1.09  &    $-$8.33  & \\ 
  VW~Cha            & 11:08:01.49   & $-$77:42:28.8    &   160  &   1.9  &   1.64    &  $-$0.78   &   1.25  &    $-$7.86 &  \\ 
  ESO~H$\alpha$~562 & 11:08:02.98   & $-$77:38:42.6    &   160  &   3.4  &   0.12    &  $-$2.01   &   0.66  &    $-$9.32 &  \\ 
  ISO$-$ChaI~143    & 11:08:22.38   & $-$77:30:27.7    &   160  &   1.3  &   0.03    &  $-$3.38   &   0.11  &    $-$10.02 &  \\ 
  Cha$-$H$\alpha-$6 & 11:08:39.52   & $-$77:34:16.7    &   160  &   0.1  &   0.07    &  $-$3.86   &   0.1   &    $-$10.25 &  \\ 
  Ass~Cha~T~2-38    & 11:08:54.64   & $-$77:02:12.9    &   160  &   1.9  &   0.13    &  $-$2.02   &   0.71  &    $-$9.35  & \\ 
  CHXR~79           & 11 09 18.13   & $-$76 30 29.2    &   160  &   5.0  &    0.25   &  $-$1.91   &   0.84  &    $-$9.55 &   \\ 
  Ass~Cha~T~2-40    & 11:09:23.79   & $-$76:23:20.8    &   160  &   1.2  &   0.55    &  $-$0.18   &   0.87  &    $-$7.30 &  \\ 
  Hn~10E            & 11 09 46.21   & $-$76 34 46.4    &   160  &   2.1  &    0.06   &  $-$2.43   &   0.36  &    $-$8.04 &   \\ 
  ISO-ChaI~207      & 11 09 47.42   & $-$77 26 29.1    &   160  &   5.0  &    0.10   &  $-$1.93   &   0.64  &    $-$9.25  &  \\ 
  Sz~32             & 11 09 53.40   & $-$76 34 25.5    &   160  &   4.3  &    0.48   &     0.061  &   1.06  &    $-$7.21  &  \\ 
  Sz~33             & 11 09 54.08   & $-$76 29 25.3    &   160  &   1.8  &    0.11   &  $-$2.10   &   0.65  &    $-$9.42  &  \\ 
  Ass~Cha~T~2-44    & 11:10:00.11   & $-$76:34:57.8    &   160  &   4.1  &   2.68    &     0.78   &   1.40  &    $-$6.6   & \\ 
  ISO$-$ChaI~237    & 11:10:11.42   & $-$76:35:29.3    &   160  &   4.1  &   0.61    &  $<-$2.47   &   1.15  &    $<-$9.79 & CH \\ 
  Sz~37             & 11 10 49.60   & $-$77 17 51.7    &   160  &   2.7  &    0.15   &  $-$0.81   &   0.51  &    $-$7.91 &   \\ 
  Ass~Cha~T~2-49    & 11:11:39.66   & $-$76:20:15.2    &   160  &   1.0  &   0.29    &  $-$0.74   &   0.36  &    $-$7.5  &  \\ 
  CHX~18N           & 11:11:46.32   & $-$76:20:09.2    &   160  &   0.8  &   1.03    &  $-$0.74   &   1.17  &    $-$8.06 &  \\ 
  Ass~Cha~T~2-51    & 11:12:24.41   & $-$76:37:06.4    &   160  &   0.1  &   0.64    &  $-$0.79   &   0.97  &    $-$8.13 &  \\ 
  Ass~Cha~T~2-52    & 11:12:27.71   & $-$76:44:22.3    &   160  &   1.0  &   2.55    &  $-$0.19   &   1.40  &    $-$7.42  & \\ 
  CW~Cha            & 11 12 30.91   & $-$76 44 24.0    &   160  &   2.1  &    0.18   &     0.85   &   0.75  &    $-$8.13 &   \\ 
\hline\\
\multicolumn{10}{c}{Chamaeleon II}\\
\hline\\
  Sz~49              & 13 00 53.23   & $-$76 54 15.2    &   178  &   0.8  &    0.11    &   $-$2.01  &   0.75  &   $-$8.94  &  \\ 
  Sz~48SW            & 13 00 53.56   & $-$77 09 08.3    &   178  &   3.7  &    0.27   &    $-$2.83  &   0.49   &  $-$9.81  &  \\ 
  Sz~50              & 13:00:55.32  & $-$77:10:22.2     &   178  &   1.8   &    0.57   &   $-$2.51  &   0.27   &  $-$8.96  &  \\
  CM~Cha             & 13 02 13.57   & $-$76 37 57.8    &   178  &   0.4  &    0.51   &    $-$1.55  &   0.77  &   $-$8.68  &  \\ 
  Hn~22              & 13 04 22.85   & $-$76 50 05.5    &   178  &   0.4  &    0.21   &    $-$2.71  &   0.38   &  $-$9.61  &  \\ 
  Hn~23              & 13 04 24.11   & $-$76 50 01.2    &   178  &   0.8  &    1.11   &    $-$1.55  &   1.01  &   $-$8.69  &  \\ 
  Hn~24              & 13:04:55.75  & $-$77:39:49.5     &   178  &   1.9   &    0.85   &   $-$1.99  &   1.00   &  $-$9.07   & \\
  Sz~53              & 13 05 12.69   & $-$77 30 52.3    &   178  &   2.8  &    0.11   &    $-$1.23  &   0.32  &   $-$8.16  &  \\ 
  Sz~58              & 13 06 57.45   & $-$77 23 41.5    &   178  &   3.5  &    0.53   &    $-$1.25  &   0.9   &   $-$8.46  &  \\ 
  Sz~61              & 13 08 06.28   & $-$77 55 05.2    &   178  &   1.6  &    1.61   &    $-$1.05  &   0.98   &  $-$8.09  &  \\ 

\hline\\[-5pt]                       
\multicolumn{10}{c}{$\sigma$ Ori}\\
\hline\\    [+5pt]
 SO397        & 05 38 13.18   & $-$02 26 08.63    &    360  &  0.0  &  0.19  &   $-$2.71   &   0.20   &   $-$9.42  & \\
 SO490        & 05 38 23.58   & $-$02 20 47.47    &    360  &  0.0  &  0.08  &   $-$3.10   &   0.14   &   $-$9.97  & \\
 SO500        & 05 38 25.41   & $-$02 42 41.15    &    360  &  0.0  &  0.02  &   $-$3.95   &   0.08   &   $-$10.27 &   \\
 SO587        & 05 38 34.04   & $-$02 36 37.33    &    360  &  0.0  &  0.28  &   $-$4      &   0.20   &   $-$10.41 &  \\
 SO646        & 05 38 39.01   & $-$02 45 31.97    &    360  &  0.0  &  0.1   &   $-$3      &   0.3    &   $-$9.68  & \\
 SO848        & 05 39 01.94   & $-$02 35 02.83    &    360  &  0.0  &  0.02  &   $-$3.5    &   0.19   &   $-$10.39 &  \\
 SO1260       & 05 39 53.63   & $-$02 33 42.88    &    360  &  0.0  &  0.13  &   $-$2      &   0.26   &   $-$8.97  & \\
 SO1266       & 05 39 54.22   & $-$02 27 32.87    &    360  &  0.0  &  0.06  &   $-$4.8    &   0.20   &   $-$11.38 &  \\
 \hline\\    [+5pt]
\end{longtable}    

 \textbf{Notes:} \\
 TD : transitional discs \\
 SL : sub-luminous objects \\
 CH : accretion luminosity compatible with chromospheric emission. \\

\newpage

\begin{landscape}
\begin{longtable}{l|ccccc|cccccc}
\caption[]{\osei\, parameters}\\
\hline
Source & \multicolumn{5}{c}{LVC} & \multicolumn{5}{c}{HVC} & \\
       & F$\pm \Delta$F & V$_p$ & FWHM & $\Delta$V\footnotemark[1] & log($L_{[\ion{O}{i}]}$)\footnotemark[2] & F$\pm \Delta$F & V$_p$ & FWHM & $\Delta$V\footnotemark[1]& log($L_{[\ion{O}{i}]}$)\footnotemark[2] &  \\
       &  10$^{-16}$\,erg\,s$^{-1}$\,cm$^{-2}$ & \kms & \kms & \kms & L$_\odot$ & erg\,s$^{-1}$\,cm$^{-2}$ & \kms & \kms & \kms & L$_\odot$ & \\
\hline
\endfirsthead
\caption{Continued.} \\
\hline
Source & \multicolumn{5}{c}{LVC} & \multicolumn{5}{c}{HVC} &\\
       & F$\pm \Delta$F & V$_p$ & FWHM 
       & $\Delta$V\footnotemark[1] & log($L_{[\ion{O}{i}]}$)\footnotemark[2]
       & F$\pm \Delta$F & V$_p$ & FWHM
       & $\Delta$V\footnotemark[1] & log($L_{[\ion{O}{i}]}$)\footnotemark[2]&  \\
       
       &  10$^{-16}$\,erg\,s$^{-1}$\,cm$^{-2}$ & \kms & \kms 
       & \kms & L$_\odot$ & erg\,s$^{-1}$\,cm$^{-2}$ & \kms
       & \kms & \kms & L$_\odot$ & \\
   \hline
\endhead
\multicolumn{11}{c}{Lupus GTO sample}\\
\hline
 Sz~66            & 225$\pm$6     &    $-$19.4  &  48.6  &  35.2   &     $-$4.81    &   $<$15             &  ...      &  ...    &  ...     &    $<-$6.463   &\\  
 AKC2006-I        & $<$0.285      &    ...    & ...    &  ...    &  $<-$7.715   &   $<$0.3                &  ...      &  ...    &  ...     &    $<-$8.162   & \\ 
 Sz~69            & 60.9$\pm$2.0  &    $-$7.7   &  55.5  &  43.2   &    $-$5.377    &   34.4$\pm$1.0      &   $-$92.9 &   73.5  &  65.2    &    $-$5.626    &  HVB\footnotemark[3]\\
                  &               &             &        &         &                &   13.4$\pm$50       &   $+$58.8 &   48.3  &  33.9    &    $-$6.03     &  HVR\footnotemark[3]\\
 Sz~71            & 30.0$\pm$1.0  &    $-$1     &  97.5  &  91.9   &    $-$5.685    &   $<$3              &  ...      &   ...   & ...      &    $<-$7.162   & \\ 
 Sz~72            & 28.0$\pm$4.0  &    $-$10.8  &  42.2  &  24.7   &    $-$5.715    &   45.7$\pm$7.0      &   $-$125. &   72.3  &  63.8    &    $-$5.502    &  HVB\\
 Sz~73            & 415$\pm$2     &    $-$18.9  &  60.8  &  50.6   &    $-$4.544    &   932$\pm$4         &   $-$94.2 &   59.1  &  48.3    &    $-$4.193    &  HVB\\
 Sz~74            & 213$\pm$25    &    $-$7.5   &  40.5  &  21.1   &    $-$4.834    &   $<$75             &  ...      &   ...   &  ...     &    $<-$5.764   & \\ 
 Sz~83            & 829$\pm$20    &    $-$19.9  & ...    &  21.1   &    $-$4.244    &   1260$\pm$20       &   $-$145. &   190   &  186.9   &    $-$4.062    &  HVB\\
 Sz~84            & 14.5$\pm$1.5  &    $-$7.4   &  41.4  &  22.9   &    $-$6.001    &   $<$4.5            &  ...      &  ...    &  ...     &    $<-$6.986   & \\ 
 Sz~130           & 21.4$\pm$3.0  &    $+$17.7   &  72.9  &  64.6   &   $-$5.832    &   47.9$\pm$6.0      &   $-$49.3 &   57.4  &  46.2    &    $-$5.482    & HVB\\
 Sz~88A           & 134$\pm$10    &    $-$1.7   &  53.5  &  40.7   &    $-$4.785    &   $<$30             &  ...      &   ...   &  ...     &    $<-$5.912   & \\ 
 Sz~88B           & 6.07$\pm$0.80 &    $-$3.9   &  55.6  &  44.5   &    $-$6.129    &   $<$2.4            &  ...      &   ...   &  ...     &    $<-$7.009   & \\ 
 Sz~91            & 40.0$\pm$4.8  &    $-$3.4   & 34     &$<$34.0  &    $-$5.31    &   $<$10              &  ...      &   ...   &  ...     &    $<-$5.435   & \\ 
 Lup713           & $<$0.43       &     ...   &  ...   &  ...    &  $<-$7.279   &   $<$0.43               &  ...      &   ...   &  ...     &    $<-$7.798   & \\ 
 Lup604s          & 1.58$\pm$0.16 &    $-$2.5   &  50.1  &  36.7   &    $-$6.714    &   $<$0.48           &  ...      &   ...   &  ...     &    $<-$7.708   & \\ 
 Sz~97            & 2.83$\pm$0.80 &    $-$36.4  &  58.4  &  47.0   &     $-$6.46    &   $<$2.4            &  ...      &   ...   &  ...     &    $<-$7.009   & \\ 
 Sz~99            & 13.6$\pm$0.9  &     $+$3.3  &  77.2  &  69.1   &    $-$5.779    &   2.2$\pm$0.6       &   $+$72.8 &   51.0  &   38.0   &    $-$6.57     &  HVR\\
 Sz~100           & 32.4$\pm$2.0  &     $-$6.6  &  43.7  &  27.9   &    $-$5.402    &    39.8$\pm$2.0     &   $-$54.6 &   84.6  &   77.5   &    $-$5.312    &  HVB\\
                  &               &             &        &         &                &   4.06$\pm$1.3      &   $+$56.2 &   50.2  &   36.9   &    $-$6.30     &  HVR\\
 Sz~103           & 37.0$\pm$3.0  &    $-$32.4  &  54.9  &  43.2   &    $-$5.344    &   $<$9              &  ...      &   ...   &  ...     &    $<-$6.435   & \\ 
 Sz~104           & 4.18$\pm$0.50 &    $-$6.8   &  40.1  &  21.1   &    $-$6.291    &   $<$1.5            &  ...      &   ...   &  ...     &    $<-$7.213   & \\ 
 Lup706           & $<$0.174      &     ...   & ...    &  ...    &  $<-$7.682   &   $<$0.17               &  ...      &   ...   &  ...     &    $<-$8.134   & \\ 
 Sz~106           & 20.0$\pm$2.0  &    $-$9.6   & 108.1  & 105.0   &    $-$5.611    &   $<$6              &  ...      &   ...   &  ...     &    $<-$6.611   & \\ 
 Par-Lup3-3       & 34.1$\pm$5.0  &    $-$1.4   &  64.8  &  55.4   &    $-$5.379    &   $<$15             &  ...      &   ...   &  ...     &    $<-$6.213   & \\ 
 Par-Lup3-4       & 46.9$\pm$5.0  &    $+$0.6   &  65.6  &  56.6   &    $-$5.241    &   $<$15             &  ...      &   ...   &  ...     &    $<-$5.736   & \\ 
 Sz~110           & 16.5$\pm$2.5  &    $-$7.1   &  66.7  &  57.7   &    $-$5.695    &   $<$7.5            &  ...      &   ...   &  ...     &    $<-$6.213   & \\ 
 Sz~111           & 80.6$\pm$8.0  &    $-$2.9   &  46.8  &  32.4   &    $-$5.006    &   $<$24             &  ...      &   ...   &  ...     &    $<-$6.009   & \\ 
 Sz~112           & 19.4$\pm$1.5  &    $-$2.4   &  49.6  &  36.7   &    $-$5.624    &   $<$4.5            &  ...      &   ...   &  ...     &    $<-$6.736   & \\ 
 Sz~113           & 6.37$\pm$0.60 &    $-$12.3  &  34    &$<$34.0  &    $-$6.108    &   21.7$\pm$1.5      &   $-$122.6&   54.5  &   42.6   &    $-$5.576    &  HVB\\  
 2MASSJ16085953   & 2.69$\pm$0.30 &    $-$16.8  &  58.3  &  47.0   &    $-$6.482    &   $<$0.9            &  ...      &   ...   &  ...     &    $<-$7.435   & \\ 
 SSTc2dJ160901.4  & 31.8$\pm$4.0  &    $-$9.9   &  74.8  &  66.8   &    $-$5.41    &   $<$12              &  ...      &   ...   &  ...     &    $<-$6.213   & \\ 
 Sz~114           & 15.9$\pm$2.0  &    $-$10.1  &  43.9  &  27.9   &    $-$5.711    &   13.2$\pm$2        &   $-$93.3 &   77.3  &   69.4   &    $-$5.792    &  HVB\\  
 Sz~115           & $<$6.0        &     ...   & ...    &  ...    &  $<-$6.134   &   $<$6                  &  ...      &   ...   &  ...     &    $<-$6.611   & \\ 
 Lup-818          & 1.30$\pm$0.17 &    $-$7.2   &  43.3  &  26.3   &    $-$6.798    &    $<$5             &  ...      &   ...   &  ...     &    $<-$6.736   & \\ 
 Sz~123A          & 69.6$\pm$4.0  &    $-$10.1  &  64.8  &  55.4   &    $-$5.07    &   23.9$\pm$1.5       &   $-$61.9 &   61.4  &   51.1   &    $-$5.534    &  HVB\\  
 Sz~123B          & 12.9$\pm$0.8  &    $-$10.6  &  66.2  &  56.6   &    $-$5.802    &   4.15$\pm$1.0      &   $-$69.6 &   65.8  &   56.3   &    $-$6.294    &  HVB\\  
 SST-Lup3-1       & $<$1.00       &     ...   & ...    &  ...    &  $<-$6.912   &   $<$1                  &  ...      &   ...   &   ...    &    $<-$7.435   & \\ 
\hline
\multicolumn{11}{c}{Lupus New Sample}\\
\hline
Sz65                            & $<$117         & ...     &  ...  &  ...  & $<-$5.09& $<$117         & ...     & ...  &  ...  & $<-$5.09  &\\            
AKC2006-18                      & 0.63$\pm$0.03  & $-$16.5 &  51.9 & 33.1  & $-$7.36 & $<$0.1        & ...     & ...  & ...   & $<-$8.15  & \\            
\footnotesize{SSTc2dJ154508.9}  & 57.3$\pm$2.5   & $+$5.0  &  50.5 & 30.8  & $-$5.40 & 108.2$\pm$2.5 & $-$92.6 & 52.2 &  33.5 & $-$5.13   & HVB\\         
Sz68                            & $<$860         & ...     &  ...  &  ...  & $<-$4.23& $<$860        & ...     & ...  &  ...  & $<-$4.23  & \\            
\footnotesize{SSTc2dJ154518.5}  & 10.28$\pm$0.04 & $-$9.9  &  47.5 & 25.6  & $-$6.15 & 3.43$\pm$0.04 & $-$53.8 & 50.9 &  31.5 & $-$6.63   & HVB\\         
                                &                &         &       &       &         & 1.12$\pm$0.04 & $-$55.9 & 50.6 &  31.5 & $-$7.11   & HVR\\         
Sz81A                           & 2.2$\pm$0.7    & $+$1.5  &  40   & ...   & $-$6.57 &               & ...     & ...  & ...   & $<-$6.57  & \\            
Sz81B                           & 6.7$\pm$0.6    & $-$12.9 &  126  & 119.5 & $-$6.09 & $<$1.8        & ...     & ...  & ...   & $<-$6.70  & \\            
Sz129                           & 171$\pm$23      & $-$1.2  &  47.5 & 25.6 & $-$4.93 & $<$28         &         &      &       & $<-$5.71  & \\            
\footnotesize{SSTc2dJ155925.2}  & 0.36$\pm$0.12  & $+$0.2  &  40   & ...   & $-$7.61 & $<$ 0.36      &...      & ...  & ...   & $<-$7.61  & \\            
RYLup                           & 391.7$\pm$94.8 & $-$1.3  &  40   & ...   & $-$4.57 & $<$284        & ...     & ..   &  ...  & $<-$4.71  & \\            
\footnotesize{SSTc2dJ160000.6}  & 1.59$\pm$0.46  & $-$9.3  &  54.6 & 37.2  & $-$6.96 & $<$1.4        & ...     & ...  & ...   & $<-$7.02  & \\            
\footnotesize{SSTc2dJ160002.4}  & $<$11.5        & ...     & ...   & ...   & $<-$6.10& $<$11.5       & ...     & ...  & ...   & $<-$6.10  & \\            
\footnotesize{SSTc2dJ160026.1}  & $<$2.2         & ...     &  ...  & ...   & $<-$6.81& $<$2.2        & ...     & ...  &  ...  & $<-$6.81  & \\            
MYLup                           & $<$85          & ...     &  ...  &  ...  & $-$5.23 & $<$85         & ...     & ...  &  ...  & $-$5.23   &\\             
Sz131                           & 28.5$\pm$2.0   & $+$2.7  &  59.4 & 43.9  & $-$5.71 & 4$\pm$2       & $-$39.5 & 68.3 & 55.4  & $-$6.59   & HVB\\         
Sz133                           & 59.5$\pm$2.9   & $-$7.3  &  75.2 & 63.7  &  $-$5.39& $<$8.6        & ...     & ...  & ...   &   ...     & \\            
\footnotesize{SSTc2dJ160703.9}  & 0.39$\pm$0.1   & $-$13.3 &  56.9 & 40.5  &   $-$7.3& $<$0.3        & ...     & ...  & ...   &   ...     & \\ 
\footnotesize{SSTc2dJ160708.6}  & 62.8$\pm$2.6   & 1.2     &  40   & ...   & $<-$5.11& 38.2$\pm$2.6  & $-$38.4 & 60.7 & 45.6  & $<-$5.33  & HVB \\         
                                &                &         &       &       &         & 9.3$\pm$2.6   & $+$30.9 & 40  & ...   & $<-$5.94  & HVR \\
Sz90                            & 110.8$\pm$5.1   & $-$1.9  &  46.5 & 23.7  & $-$4.87 & 55.3$\pm$5.1 & $-$62.8 & 45.5 & 20.7  & $-$5.17   & HVB\\         
Sz95                            & $<$9.8         &...      & ...   & ...   & $<-$5.92& $<$9.8        & ...     & ...  & ...   & $<-$5.92  & \\            
Sz96                            & 277.5$\pm$6.5  & $-$16.4 &  85.7 & 75.8  & $-$4.47 & $<$20         & ...     & ...  & ...   & $<-$5.62  & \\            
\footnotesize{2MASSJ16081497}   & 1.2$\pm$0.1  & $-$24.8 &  77.9 & 66.8  & $-$6.61 & $<$0.2        & ...     & ...  & ...   & $<-$7.46  & \\              
Sz98                            & 627.7$\pm$47.2 & $-$0.2  &  47.0 & 20.6  & $-$4.11 & 222.4$\pm$47.2& $-$70.5 & 61.2 & 46.3  & $-$4.39   & HVB\\         
Lup607                          & 0.3$\pm$0.07   & $-$11.9 &  40   & ...   & $-$7.41 & $<$0.2        & ...     & ...  & ...   & $<-$7.61  & \\            
Sz102                           & 1516$\pm$0.4   & $-$25.6 &  86.0 & 76.1  & $-$3.73 & 656.1$\pm$0.4 & $+$56.5 & 86.4 & 76.6  & $-$4.29          & HVR\\         
\footnotesize{SSTc2dJ160830.7}  & $<$335        & ...     &  ...  &  ...  & $<-$4.39& $<$335       & ...     & ...  &  ...  & $<-$4.39  & \\              
V1094Sco                        & $<$170        & ...     &  ...  &  ...  & $<-$4.68&  $<$170        & ...     &  ...  &  ...  & $<-$4.68& \\ 
Sz108B                          & 8.6$\pm$0.5    & $+$9.5  &  52.8 & 34.4  & $-$5.97 & 14.9$\pm$0.5  & $-$34.9 & 46.9 & 24.5  & $-$5.74   & HVB\\         
\footnotesize{2MASSJ16085324}   & $<$12.6         &...      & ...   & ...   & $<-$5.81& $<$12.6        & ...     & ...  & ...   & $<-$5.81  & \\          
\footnotesize{2MASSJ16085373}   & $<$7.5         & ...     &  ...  &  ...  &  ...    & ...           & ...     & ...  & ...   &  ...      &\\             
\footnotesize{2MASSJ16085529}   & 1.4$\pm$0.1    & $-$6.7  &  40   & ...   & $-$6.75 & 0.7$\pm$0.1   & $-$49.5 & 37.4 & ...   & $-$6.91   & HVB\\         
                                &                &         &       &       &         & 1.0$\pm$0.1   & $+$30   & 61.2 & 46.3  & $-$6.76   & HVR\\         
\footnotesize{SSTc2dJ160927.0}  & 46.2$\pm$0.6   & $-$10.9 &  47.9 & 26.3  & $-$5.25 & 144.5$\pm$0.6  & $-$72.2 & 50.8 & 31.3  & $-$4.75   & HVB\\        
                                &                &         &       &       &         & 31.9$\pm$0.6   & $+$59.6 & 102.8& 94.7  & $-$5.40   & HVR\\         
Sz117                           & 14.5$\pm$2.3    & $-$5.5  &  64.5 & 50.6  & $-$5.75 & 7.2$\pm$2.3          & $-$84.8    & 29  & ...   & $-$6.39  & HVB\\           
Sz118                           & 265.1$\pm$19.3   & $-$27.6 &  42.7 & 14.9  & $-$4.49 & $<$57         & ...     & ...  & ...   & $<-$5.15  & \\          
\footnotesize{2MASSJ16100133}   & 6.2$\pm$2.3    & $-$20.2 &  82.1 & 71.7  & $-$6.12 & $<$7       &...      & ...  & ...   & $<-$6.07  &\\                
\footnotesize{SSTc2dJ161018.6}  & $<$0.7         & ...     &  ...  &  ...  & $<-$7.09& $<$0.7        & ...     & ...  &  ...  & $<-$7.09  & \\            
\footnotesize{SSTc2dJ161019.8}  & $<$0.3         & ...     &  ...  &  ...  & $<-$7.48& $<$0.3        & ...     & ...  &  ...  & $<-$7.48  & \\            
\footnotesize{SSTc2dJ161029.6}  & 4.4$\pm$0.4    & $-$9.4  &  72.2 &  60.1 & $-$6.26 & $<$1.2        & ...     & ...  &  ...  & $<-$6.83  & \\            
\footnotesize{SSTc2dJ161243.8}  & 174.0$\pm$3.2   & $-$26.5 &  95.6 & 86.8  & $-$4.67 & $<$9.6         & ...     & ...  & ...   & $<-$5.93  & \\          
\footnotesize{SSTc2dJ161344.1}  & 4.33$\pm$0.21  & $+$15.7 &  53.5 & 35.5  & $-$6.27 & 1.84$\pm$0.21 & $-$29.1 & 40   & ...   & $-$6.65   & HVB\\         
\hline 
\multicolumn{11}{c}{Lupus archive}\\
\hline     
Sz75/GQ Lup                     & 350$\pm$40    & $-$3.0    &  57.0 & 54.5  & $-$4.36 &  $<$120       & ...     &  ...  &  ...  & $<-$4.83& \\ 
Sz76                            & 66$\pm$3.2    & $-$5.4    &  60.2 & 57.9  &  $-$5.09 & $<$9.6       & ...     &  ...  &  ...  & $<-$5.93& \\ 
Sz77                            & $<$90         & ...     &  ...  &  ...  & $-$4.96    & $<$90         & ...     &  ...  &  ...  & $<-$4.96 \\
RXJ1556.1-3655                  & 220$\pm$97    & $-$2.4  &  52.2 & 49.5  & $-$4.57    & $<$291         & ...     &  ...  &  ...  & $<-$4.45 \\
Sz82/IM Lup                     & $<$100        & ...     &  ...  &  ...  & $<-$4.91   & $<$100        & ...     &  ...  &  ...  & $<-$4.91 \\
EX Lup                          & $<$120        & ...     &  ...  &  ...  & $<-$4.83   & $<$120        & ...     &  ...  &  ...  & $<-$4.83 \\
\hline  
\multicolumn{11}{c}{Cha I}\\
\hline                                  
  Ass~Cha~T~2-3      & 697.1$\pm$5.5   &  $-$6     &  48.5    &  45.4   &    $-$4.26          & 121.5$\pm$5.5    & $-$67.5    &  49.0   & 45.9   &  $-$5.02   & HVB \\
                     &                 &           &          &         &                     &  136.6$\pm$5.5   & $+$48.2    &  53.3   & 50.5   &  $-$4.97  & HVR\\
  TW~Cha             & 146.4$\pm$6.9   &  $-$1.76  &   49.5   &  46.5   &    $-$4.94          &  $<$20.7           &  ...     &  ...    &  ...   &  $<-$5.79   &  \\
  CR~Cha             & 433.3$\pm$57.7  &  $+$2.5   &   29.5   &  24.1   &    $-$4.47          &  $<$173.3          &  ...     &  ...    &  ...   &  $<-$4.87   &  \\
  Ass~Cha~T~2-12     & 2.0$\pm$0.7     &  $+$8.4   &   39.5   &  35.6   &    $-$6.80          & 8.1$\pm$0.7      & $+$62.9    &  39.5   & 35.6    &  $-$6.20   & HVR \\
  CT~Cha~A           & 1045$\pm$447    &  $-$13.5  &   81.4   &  79.6   &    $-$4.09          & 856.5$\pm$44.7   & $-$114.0   &  88.6   & 86.9    &  $-$4.17   & HVB \\
  Hn~5               & 4.0$\pm$0.2     &  $-$5.1   &   33.2   &  28.5   &    $-$6.5           & $<$0.6             &  ...     &  ...    &  ...   &  $<-$7.31   &  \\
  Ass~Cha~T~2-23     & 100.9$\pm$2.0   &  $-$7.0   &   61.9   &  59.5   &    $-$5.10          & 77.5$\pm$2.0     & $-$58.6    &  33.2   & 28.52   &  $-$5.22   & HVB \\
  Ass~Cha~T~2-24     & $<$25           &  ...      &   ...    &  ...    &    $<-$5.71          & $<$25              &  ...     &  ...    &  ...   &  $<-$5.71   &  \\
  Cha$-$H$\alpha-$1  &  1.3$\pm$0.6    &  $-$13.7  &    48    &  44.9   &    $-$7.00          & $<$1.8             &  ...     &  ...    &  ...   &  $<-$7.71   &  \\
  Cha$-$H$\alpha-$9  &  $<$11.6        &  ...      &   ...    &  ...    &    $<-$6.04          & $<$11.6            &  ...     &  ...    &  ...   &  $<-$6.04   &  \\
  Sz~19              &  $<$422.4       &  ...      &   ...    &  ...    &    $<-$4.24          &  $<$422.4          &  ...     &  ...    &  ...   &  $<-$4.48   &  \\
  Sz~22              &  977.6$\pm$5.8  &  $+$0.7   &   30.9   &  35.1   &    $-$4.12          &  853.6$\pm$5.8   & $-$92.0    &  172.4  &  171.6  &  $-$4.17   & HVB \\
                     &                 &           &          &         &                     &  745.2$\pm$5.8   & $+$71.5    &  112.4  &  111.1  &  $-$4.23   & HVR \\
  VW~Cha             &  146.4$\pm$6.9  &  $+$13.5  &  66.2    &  63.9   &    $-$3.85          & $<$72              &  ...     &  ...    &  ...   &  $<-$5.25   &  \\
  ESO~H$\alpha$~562  &  823.5$\pm$3.0  &  $-$6.8   &  45.2    & 41.9    &    $-$4.19          & 1882.6$\pm$3.0   & $-$71.8    &   47.1  &  43.9   &  $-$3.83   & HVB \\
                     &                 &           &          &         &                     & 386.8$\pm$3.0    & $+$104.6   &   67.6  &  67.4   &  $-$4.52   & HVR \\
  ISO$-$ChaI~143     &  21.0$\pm$0.1   &  $+$13.8  &   81.4   &  79.6   &    $-$5.78          & $<$0.4             &  ...     &  ...    &  ...   &  $<-$7.56   &  \\
  Cha$-$H$\alpha-$6  &   $<$0.25       &  ...      &   ...    &  ...    &   $<-$7.71          & $<$0.25           &  ...     &  ...    &  ...   &  $<-$7.71   &  \\
  Ass~Cha~T~2-38     &  443.2$\pm$3.1  &  $+$0.6   &  69.5    &  67.4   &    $-$4.46          & $<$9.4             &  ...     &  ...    &  ...   &  $<-$6.13   &  \\
  CHXR~79            &  102.8$\pm$8.6  &  $-$23    &  47      &  43.8   &    $-$4.84          &  114.3$\pm$8.6   &  $-$78.7   &   42.7  &  39.17  &  $-$5.05   & HVB \\
                     &                 &           &          &         &                     & 41.7$\pm$8.6     &  $+$32     &   24    &  16.9   &  $-$5.48   & HVR \\        
  Ass~Cha~T~2-40     & 190.5$\pm$5.9   &  0.0      &  38      &  33.9   &    $-$4.83          & $<$21              &  ...     &  ...    &  ...   &  $<-$5.78   &  \\
  Hn~10E             & $<$5.7E-16      &  ...      &   ...    &  ...    &    $<-$6.06          & $<$5.7E-16         &  ...     &  ...    &  ...   &  $<-$6.35   &  \\
  ISO-ChaI~207       &   56.7$\pm$2.4  &  $-$0.9   &  14.2    &  ...    &    $-$4.74          & $<$7.3             &  ...     &  ...    &  ...   &  $<-$6.24   &  \\
  Sz~32              & 2267.9$\pm$12.3 &  $-$13.3  &   47     &  43.8   &    $-$3.72          & 2399.3$\pm$12.3  &  $-$79.3   &   142   & 140.979 &  $-$3.73   & HVB \\
  Sz~33              &  27.1$\pm$1.0   &  $-$8.0   & 34       & 29.4    &    $-$4.78          & 10.3$\pm$1.0     &  $-$55     &   47.4  &  44.24  &  $-$6.09   & HVB \\
  Ass~Cha~T~2-44     &  3776$\pm$187   &  $-$5.2   &  68.6    & 66.5    &    $-$3.53          & $<$563             &  ...     &  ...    &  ...   &  $<-$4.36   &  \\
  ISO$-$ChaI~237     &  $<$30.7        &  ...      &   ...    &  ...    &    $<-$5.62          & $<$30.7            &  ...     &  ...    &  ...   &  $<-$5.62   &  \\
  Sz~37              & 71.3$\pm$3.4    &  $-$5.0   &  46      & 42.7    &    $-$4.98          & $<$10.3            &  ...     &  ...    &  ...   &  $<-$6.09   &  \\
  Ass~Cha~T~2-49     &  277.4$\pm$3.1  &  $-$6.2   &  39.0    & 35.1    &    $-$4.66          & 292.8$\pm$3.1    &  $-$78.0   &  47.4    &  44.2  &  $-$4.64   &  HVB\\
  CHX~18N            &  380.2$\pm$25.1 &  $-$8.0   &  63.8    & 61.5    &    $-$4.53          &  $<$75.3           &  ...     &  ...    &  ...   &  $<-$5.23   &  \\
  Ass~Cha~T~2-51     & $<$53.0         &  ...      &   ...    &  ...    &    $<-$5.38          &  $<$53.0           &  ...     &  ...    &  ...   &  $<-$5.38   &  \\
  Ass~Cha~T~2-52     &  615.2$\pm$47.1 &  $+$2.58  &   38     & 33.9    &    $-$4.32          & $<$125.7           &  ...     &  ...    &  ...   &  $<-$5.01   &  \\
  CW~Cha             & 349.5$\pm$2.5   &  $+$0.8   &   47     & 43.8    &    $-$4.47          & 90.9$\pm$2.5     &  $-$99.6   &  85     &  83.28  &  $-$5.15   & HVB \\
                     &                 &           &          &         &                     & 115.8$\pm$2.5    &  $+$86.6   &  28.5   &  22.9   &  $-$5.04   & HVR \\
\multicolumn{11}{c}{Cha II}\\
\hline 
  Sz~49              & 59.2$\pm$1.7    &  $+$0.4  &  28.4  &  22.7      &    $-$5.24         &  28.9$\pm$1.7    & $+$139.5  &   85    &    83.28 & $-$5.55    & HVR \\
  Sz~48SW            & 81.5$\pm$3.7    &  $+$16.5 &  76    &   74.07    &    $-$5.10         &  $<$30             &  ...     &  ...    &  ...   & $<-$5.54    &  \\
  Sz~50              & $<$17.0         &  ...      &   ...    &  ...    &    $<-$5.87         &  $<$17.0           &  ...     &  ...    &  ...   & $<-$5.87    &  \\
  CM~Cha             & 101.6$\pm$9.3   & $-$10    &  47    &   43.8     &    $-$5.01         &  $<$28.0           &  ...     &  ...    &  ...   & $<-$5.57    &  \\
  Hn~22              & $<$3            &  ...      &   ...    &  ...    &    $<-$6.54         &  $<$3            &  ...     &  ...    &  ...   & $<-$6.54    &  \\
  Hn~23              & 328.1$\pm$18.9  & $-$5.4   &  57    &  54.4      &    $-$4.49         &  $<$44             &  ...     &  ...    &  ...   & $<-$5.37    &  \\
  Hn~24              & 104.9$\pm$10.0  & $-$6     &  52.4  &  49.6      &    $-$5.08         & $<$20              &  ...     &  ...    &  ...   & $<-$5.80    &  \\
  Sz~53              & 373.9$\pm$5.9   & $-$18.5  &  71    &  68.9      &    $-$4.44         &  164.5$\pm$5.9   &   $-$58.1 &   38    &    33.98 & $-$4.80    & HVB \\ 
  Sz~58              & 331.9$\pm$11.5  & $-$37.5  &  95    &  93.467    &    $-$4.49         &  $<$54             &  ...     &  ...    &  ...   & $<-$5.28    &  \\
  Sz~61              & 633.6$\pm$23.8  & $-$4.3   &  47    &  43.8      &    $-$4.21         & 489.2$\pm$23.8   &   $-$129  &  157    &   156.07 & $-$4.32    & HVB \\ 
\hline
\multicolumn{11}{c}{\sig Ori}\\
\hline 
SO397               & $<$1.00        &   ...      & ...    &   0    &  $<-$6.402 &  $<$1          &  ...    &  ...   &  ...   &  $<-$6.8     & \\
SO490               & 0.51$\pm$0.1   &   $-$22.7  &  62.8  &  52.8  &  $-$6.69   &  $<$0.3        &  ...    &  ...   &  ...   &  $<-$7.402   & \\
SO500               & 0.24$\pm$0.07  &   $-$21.2  &  80.9  &  73.4  &  $-$7.021  &  0.04$\pm$0.01 &  $-$70.3  &  34    &  $<$34 &  $-$7.854      & HVB \\
SO587               & 14.3$\pm$0.7   &    $-$7.5  &  56.0  &  44.5  &  $-$5.246  &  $<$2          &   ...   &  ...   &  ...   &  $<-$6.557   &\\ 
SO646               & 6.00$\pm$0.60  &    $-$9.3  &  71.8  &  63.2  &  $-$5.624  &  $<$18         &  ...    &  ...   &  ...   &  $<-$5.624   & \\
SO848               & 4.61$\pm$0.15  &    $-$8.5  &  42.2  &  25.1  &  $-$5.738  &  3.6$\pm$0.1   &  $-$51.8  &   48.6 &  34.7  &  $-$5.845      & HVB \\
SO1260              & 4.20$\pm$0.35  &    $-$6.6  &  41.7  &  24.1  &  $-$5.778  &  $<$10         &  ...    &  ...   &  ...   &  $<-$5.925   & \\
SO1266              & 1.88$\pm$0.20  &    $-$7.5  &  49.2  &  35.6  &  $-$6.128  &  $<$6          & ...     &  ...   &  ...   &  $<-$6.101   & \\
\hline                                                    
\end{longtable}
\footnotetext[1]{Line width deconvolved for instrumental profile}
\footnotetext[2]{Dereddened line luminosity}
\footnotetext[3]{HVB and HVR refer to High Velocity Blue and Red components, respectively}

\end{landscape}
 
\section{Kinematical information on the \sii\ and \nii\ lines.}

Table C1 summarizes the kinematical parameters (V$_p$ and FWHM)
derived from Gaussian fitting of the  \sii\ 6730.8\AA\ and \nii\ 6583.4\AA\ lines. 
The corresponding parameters of the \osei\ line, already listed in Table B2, are also added here for comparison. 

\begin{footnotesize}
\begin{table*}
\label{tab_b1}
\caption[]{Kinematical parameters of the \sii\ and \nii\ lines}
\vspace{0.5cm}
\begin{tabular}{lrrrrrrr}
\hline
Source &  \multicolumn{2}{c}{[\ion{Ni}{ii}]6730.8\AA} & \multicolumn{2}{c}{[\ion{S}{ii}]6583.4\AA} & \multicolumn{2}{c}{\osei}\\
\\
       &  \multicolumn{1}{c}{V$_p$}   & \multicolumn{1}{c}{FWHM}   &  \multicolumn{1}{c}{V$_p$} & \multicolumn{1}{c}{FWHM} & 
       \multicolumn{1}{c}{V$_p$}   &  \multicolumn{1}{c}{FWHM}   &Comp.\\
       &   \kms   & \kms   &  \kms   & \kms   & \kms   & \kms   &\\
\\
\hline\\
Sz69             &          &           &           &           &       -7.7    &  55.5 & LVC\\
                 &  -90.3   &   62.7    &  -103.2   &   51.6    &       -92.9   &  73.5 & HVB\\
                 &   57.0   &   44.3    &    62.2   &   37.5    &        58.8   &  48.3 & HVR\\
Sz72             &          &           &           &           &       -10.8   &  42.2 & LVC\\
                 &          &           &  -132.2   &   40.8    &       -125.0  &  72.3 & HVB\\
Sz73             &          &           &           &           &        -18.9  &  60.8 & LVC\\
                 &   -94.5  &  103.0    &  -98.3    &   64.9    &        -94.2  &  59.1 & HVB\\
Sz130            &          &           &           &           &         17.7  & 72.9 & LVC\\
                 &   -56.5  &   40.8    &           &           &       -49.3   &  57.4 &  HVB\\
Sz99             &          &           &           &           &         3.3   &  77.2 & LVC\\
                 &          &           &    61.9   &   80.7    &        72.8   &  51.0 & HVR\\
Sz100            &          &           &   -12.3   &    52.3   &        -6.6   &  43.7 & LVC\\
                 &   -70.8  &   34.9    &   -71.2   &   61.8    &        -54.6  &  84.6 &  HVB\\
                            &          &           &           &          56.2  &  50.2 & HVR\\        
Sz113            &          &           &           &           &       -12.3   &  34  & LVC\\
                 &   -125.1 &   40.6    &   -120.9  &   65.5    &       -122.6  &  54.6 &  HVB\\
Sz123A           &          &           &           &           &       -20.1   &  64.8 & LVC\\
                 &    -55.8 &   38.9    &           &           &       -61.9   &  61.4 &  HVB\\
Sz123B           &          &           &           &           &        -10.6 &   66.2 & LVC\\
                 &    -56.6 &   44.4    &           &           &        -69.6  &  65.8 &  HVB\\
SSTc2dJ154508.9  &          &           &            &            &     $+$5.0    &  50.5 &LVC\\
                 &  -81.05  &   35.04   &    -70.06  &    52.82   &     $-$92.6   & 52.2 &  HVB\\
SSTc2dJ154518.5  &   -18.9  &   46.72   &            &            &     $-$9.9    &  47.5&LVC\\
                 &          &           &            &            &     $-$53.8   & 50.9& HVB\\
SSTc2dJ160708.6  &   -21.6  &   53.1    &      0.4   &    66.8    &     $+$1.2    &  40&LVC\\
                 &   -80.6  &   34.1    &    -35.2   &    45      &     $-$38.4   & 60.7& HVB\\
                 &    31.9  &   57.8    &            &            &     $+$30.9   & 40 & HVR\\
Sz102            &   -6.9   &  139.7    &      7.6   &    79.3    &               &      & LVC\\
                 & -164.7   &   71.5    &    -25.0   &    74.7    &     $-$25.6   &  86.0& HVB\\
                 &   64.0   &   71.8    &     67.0   &    77.8    &     $+$56.5   & 86.4& HVR\\
SSTc2dJ160927.0  &    -4.6  &   27.1    &      9.3   &    36.7    &     $-$10.9   &  47.9& LVC\\
                 &   -57.6  &   27.4    &    -27.1   &    38.5    &     $-$72.2   & 50.8& HVB\\
                 &          &           &            &            &     $+$59.6   & 102.8 & HVR \\                       
CT-Cha-A         &          &           &    -25.3  &     70.0   &     $-$13.5   &   81.4& LVC\\
                 &          &           &    -109.3 &     73.00  &     $-$114.0  &  88.6& HVR\\
Sz22             &          &           &     7.3    &    33.9    &     $+$0.7    &   30.9& LVC\\
                 &          &           &   -69.3    &   201.1    &     $-$92.0   &  172.4& HVB\\
                 &          &           &    89.3    &    57.1    &     $+$71.5   &  112.4& HVR\\
ESO-Ha-562       &          &    45.5   &     +4.6   &     48.0   &      $-$6.8   &  45.2& LVC\\
                 &  -62.0   &   56.3    &    -64.1   &    49.0    &     $-$71.8   &   47.1& HVB\\
                 &  127.5   &   48.1    &    147.0   &    45.3    &     $+$104.6  &   67.6& HVR\\
Sz32             &  -17.5   &   83.1    &    -6.1    &    47.6    &     $-$13.3   &   47& LVC\\
                 &          &           &    -63.9   &    100.6   &     $-$79.3   &   142& HVB\\
Ass-Cha-T-2-49   &          &           &            &    32.7    &     $-$6.2    &  39.0 & LVC\\
                 &  -54.1   &   33.1    &    -51.4   &    32.7    &     $-$78.0   &  47.4& HVB\\
Sz53             &          &           &    -23.9   &    24.8    &     $-$18.5   &   71 & LVC\\
                 &  -45.0   &   37.8    &            &            &     $-$58.1   &   38 & HVB\\
SO848            &          &           &    -3.0    &    50.0    &     -8.5      &   42.2& LVC\\
                 &  -47.9   &   50.7    &    -47.3   &    49.6    &     -51.8     &   48.2& HVB\\
\hline\\[-5pt]
\end{tabular}
\end{table*} 
\end{footnotesize}

\section{PV diagrams of \osei\ emission}

The X-shooter 2D spectral images were inspected to detect any shift of the \osei\ emission
with respect to the continuum, confirming the origin of the HVC from an extended jet
component. We found a displacement of the HVC peak emission in five bright sources.
The corresponding PV diagrams are shown in Fig. C.1

\begin{figure}[h]
\includegraphics[angle=0,width=10cm]{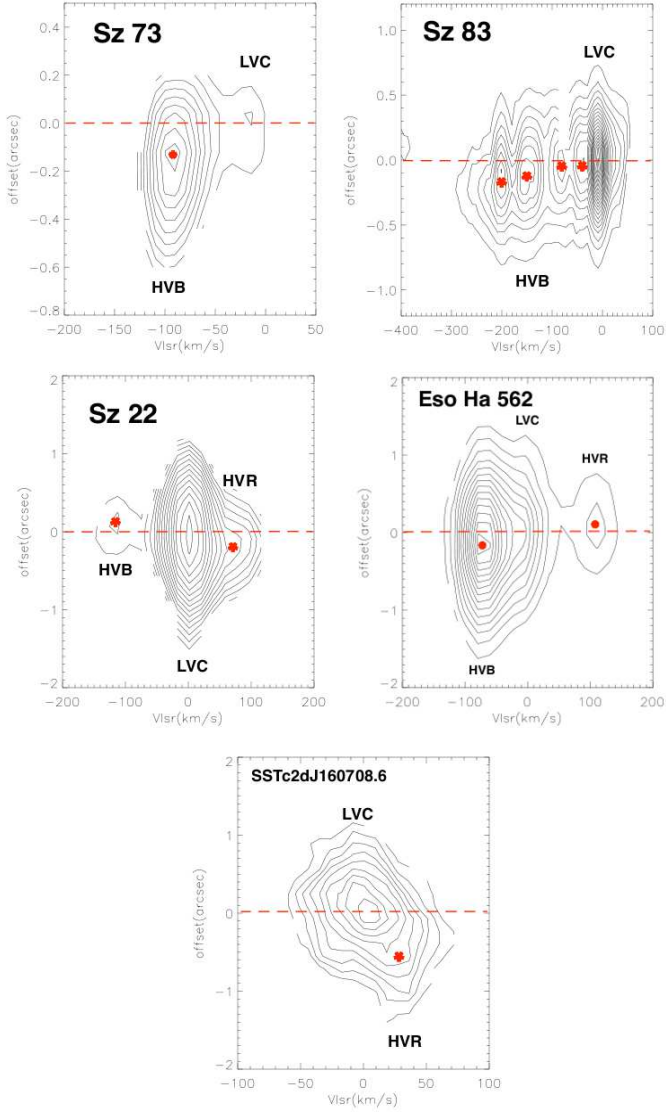}
     \caption{\osei\ continuum subtracted position velocity (PV) diagrams of five sources
     exhibiting strong HVC. The straight line indicates the position of the continuum while red dots
     highlight the peaks of the blue- and red-shifted HVC (i.e. HVB and HVR, respectively) and of the LVC. 
     The HVC peaks are systematically shifted with respect
     to the continuum position, showing that this component is extended. 
     }
         \label{PV}
\end{figure}

\end{appendix}

\end{document}